\documentclass[12pt]{article}

\usepackage[makeroom]{cancel}%Added by Martin, to cross out parts of equations with the command \cancel{}
\usepackage[normalem]{ulem}
\usepackage{amsmath,amssymb}
\usepackage{slashed}
\usepackage{xcolor} %Define colors
\usepackage{graphicx}
\usepackage{cite}
\usepackage{pdflscape}
\usepackage{multirow}
\usepackage[thinlines]{easytable}
\usepackage{url}

\newcommand{\eref}[1]{Eq.~\eqref{eq:#1}}

\newcommand{\aref}[1]{Appendix~\ref{app:#1}}
\newcommand{\sref}[1]{Section~\ref{sec:#1}}
\newcommand{\cref}[1]{Chapter~\ref{ch:.#1}}
\newcommand{\tref}[1]{Table~\ref{tab:#1}}

\newcommand{\nn}{\nonumber \\}  
\newcommand{\nnl}{\nonumber \\}

\newcommand{\beq}{\begin{equation}} 
\newcommand{\eeq}{\end{equation}} 
\newcommand{\ba}{\begin{array}}  
\newcommand{\ea}{\end{array}} 
\newcommand{\bea}{\begin{eqnarray}}  
\newcommand{\eea}{\end{eqnarray} }  
\newcommand{\be}{\begin{eqnarray}}  
\newcommand{\ee}{\end{eqnarray} }  
\newcommand{\bal}{\begin{align}}
\newcommand{\eal}{\end{align}}   
\newcommand{\bi}{\begin{itemize}}  
\newcommand{\ei}{\end{itemize}}  
\newcommand{\ben}{\begin{enumerate}}  
\newcommand{\een}{\end{enumerate}}  
\newcommand{\bc}{\begin{center}}
\newcommand{\ec}{\end{center}} 
\newcommand{\bt}{\begin{table}}
\newcommand{\et}{\end{table}}  
\newcommand{\btb}{\begin{tabular}}
\newcommand{\etb}{\end{tabular}}  
\newcommand{\bvec}{\left ( \ba{c}}
\newcommand{\evec}{\ea \right )}

\newcommand{\cO}{{\mathcal O}} 
 
\newcommand{\cL}{{\mathcal L}}

\newcommand{\re}{{\mathrm{Re}} \,}

\def\hc{{\rm h.c.}} 
\def\ov{\overline}  
  
\newcommand{\eps}{\epsilon}
\newcommand{\eL}{\bar{\epsilon}_L}

\newcommand{\eS}{\epsilon_S}
\newcommand{\eP}{\epsilon_P}
\newcommand{\eT}{\epsilon_T}

\newcommand{\bfblue}[1]{\color{blue} \mathbf{#1}}

\begin{document}

\begin{titlepage}

\vspace*{-2cm}
\begin{flushright}
LPT Orsay 17-29 \\
\vspace*{2mm}
%\today
\end{flushright}

\begin{center}
\vspace*{15mm}

\vspace{1cm}
{\LARGE \bf
Compilation of low-energy constraints  \\ 
on 4-fermion operators in the SMEFT 
} 
\vspace{1.4cm}

\renewcommand{\thefootnote}{\fnsymbol{footnote}}
{Adam~Falkowski$^a$, Mart\'{i}n~Gonz\'{a}lez-Alonso$^{b,c}$, Kin~Mimouni$^d$}
\renewcommand{\thefootnote}{\arabic{footnote}}
\setcounter{footnote}{0}

\vspace*{.5cm}
\centerline{$^a${\it Laboratoire de Physique Th\'{e}orique, CNRS, Univ. Paris-Sud,}}
\centerline{{\it Universit\'{e}  Paris-Saclay, 91405 Orsay, France}}
\centerline{$^b${\it
IPN de Lyon/CNRS, Universite Lyon 1, Villeurbanne, France}}
\centerline{${}^c$ \it CERN, Theoretical Physics Department, Geneva, Switzerland}
\centerline{${}^d$ \it Institut de Th\'{e}orie des Ph\'{e}nom\`{e}nes Physiques, EPFL, Lausanne, Switzerland }

\vspace*{.2cm}

\end{center}

\vspace*{10mm}
\begin{abstract}\noindent\normalsize

We compile information from low-energy observables sensitive to flavor-conserving 4-fermion operators with two or four leptons. 
Our analysis includes data from $e^+e^-$ colliders, neutrino scattering on electron or nucleon targets, atomic parity violation, parity-violating electron scattering, and the decay of pions, neutrons, nuclei and tau leptons. We recast these data as tree-level constraints on 4-fermion operators in the Standard Model Effective Field Theory (SMEFT) where the SM Lagrangian is extended by dimension-6 operators.
 We allow all independent dimension-6 operators to be simultaneously present with an arbitrary flavor structure. 
 The results are presented as a multi-dimensional likelihood function in the space of dimension-6 Wilson coefficients, which retains information about the correlations. 
 In this form, the results can be readily used to place limits on masses and couplings in a large class of new physics theories.

\end{abstract}

\end{titlepage}
\newpage 

\renewcommand{\theequation}{\arabic{section}.\arabic{equation}} 
%%%%%%%%%%%%%%%%%%%%%%%%%%%%%%%%%%%%%%%%%%%%%%%

 %%%%%%%%%%%%%%%%%%%%%%%%%%%%%%%%% 
\section{Introduction}

The ongoing exploration of the high-energy frontier at the LHC strongly suggests that the only fundamental degrees of freedom at the weak scale are the Standard Model (SM) ones. 
Moreover, their perturbative interactions are well described by the most general renormalizable SM Lagrangian invariant under the $SU(3) \times SU(2) \times U(1)$ local symmetry.  
A large number of precision measurements has been performed in order to test the SM predictions. 
The motivation is that some unknown heavy particles may affect the coupling strength or induce new effective interactions between the SM particles.   

One framework designed to describe  such effects in a systematic fashion goes under the name of the SM Effective Field Theory (SMEFT). 
In this approach, the SM particle content and symmetry structure is retained, but the usual renormalizability requirement is abandoned such that interaction terms with canonical dimensions $D>4$ are allowed in the Lagrangian.  
These higher-dimensional operators encode, in a model-independent way, the effects of new particles with masses above the weak scale. 
One can then analyze experimental searches once and for all within this framework. The output of such analysis, namely numerical values for the Wilson coefficients of higher-dimensional operators, can then be applied to any new physics model covered by the SMEFT. Significant progress has been recently achieved concerning the automation of this EFT matching~\cite{Henning:2014wua,delAguila:2016zcb,Ellis:2016enq,Henning:2016lyp,Fuentes-Martin:2016uol,Zhang:2016pja}. The efficient SMEFT program should be compared with model-dependent studies where non-trivial hadronic effects, PDFs, radiative corrections, experimental errors, cuts, etc., have to be taken into account for each model.

Assuming lepton number conservation, leading SMEFT contributions are expected to originate from dimension-6 operators~\cite{Leung:1984ni,Buchmuller:1985jz}.
There is a vigorous program to characterize the effects of the dimension-6 operators on precision observables and derive constraints on their Wilson coefficients in the  SMEFT Lagrangian~\cite{Han:2004az,Han:2005pr,Barbieri:2004qk,Grojean:2006nn,Cacciapaglia:2006pk,Cirigliano:2009wk,Carpentier:2010ue,Filipuzzi:2012mg,Blas:2013ana,Pomarol:2013zra,Elias-Miro:2013mua,Dumont:2013wma,Chen:2013kfa,deBlas:2013qqa,Willenbrock:2014bja,Gupta:2014rxa,Masso:2014xra,deBlas:2014ula,Ciuchini:2014dea,Ellis:2014jta, Falkowski:2014tna,delAguila:2014soa,Berthier:2015oma,Corbett:2015ksa,Efrati:2015eaa,Gonzalez-Alonso:2015bha,Buckley:2015nca,deBlas:2015aea,Falkowski:2015fla, delAguila:2015vza,Wells:2015eba,Berthier:2015gja,Falkowski:2015jaa,Ellis:2015sca,Englert:2015hrx,Falkowski:2015krw,deBlas:2016ojx,Ciuchini:2016sjh,Berthier:2016tkq,Bjorn:2016zlr, Hartmann:2016pil}.  
Most of these analyses assume that the dimension-6 operators respect some flavor symmetry in order to reduce the number of independent parameters.  
On the other hand,  Refs.~\cite{Efrati:2015eaa,Falkowski:2015krw} allowed for a completely general set of dimension-6 operators,  demonstrating that the more general approach is feasible.   
 
This paper further pursues the approach of Refs.~\cite{Efrati:2015eaa,Falkowski:2015krw}, providing new constraints on the SMEFT where all independent dimension-6 operators may be simultaneously present with an arbitrary flavor  structure. 
We compile information from a plethora of low-energy flavor-conserving experiments sensitive to electroweak gauge boson interactions with fermions and to 4-fermion operators with 2 leptons and 2 quarks (LLQQ) and 4 leptons (LLLL).   
There are two main novelties compared to the existing literature. 
First, precision constraints on the LLQQ operators  have not been attempted previously in the flavor-generic situation.  
Therefore our results are relevant to a larger class of UV completions where new physics couples with a different strength to the SM generations. 
Note that, in particular, all models addressing the recent B-meson anomalies (see e.g.~\cite{Descotes-Genon:2013wba,Altmannshofer:2013foa,Jager:2014rwa,Freytsis:2015qca,Altmannshofer:2017fio}) must necessarily involve exotic particles with flavor non-universal couplings to quarks and leptons. 
Our analysis provides model-independent constraints that have to be satisfied by all such constructions.  
Second, we include in our analysis the low-energy flavor observables (nuclear, baryon and meson decays) recently summarized in  Ref.~\cite{Gonzalez-Alonso:2016etj}. 
At the parton level these processes are mediated by the quark transitions $d(s)\to u \ell \bar{\nu}_\ell$, hence they can probe the LLQQ operators. 
We will show that for certain operators the sensitivity of these observables is excellent, such that new stringent constraints can be obtained.    
Moreover, the low-energy flavor observables offer a sensitive probe of  the W boson couplings to right-handed quarks. 
 
Our analysis is performed at the leading order in the SMEFT. 
We ignore the effects of dimension-6 operators suppressed by a loop factor, except for the renormalization group running within a small subset of the LLQQ operators.  
Moreover all dimension-8 and higher operators are neglected, and only the linear contributions of the dimension-6 Wilson coefficients are taken into account. 
The corollary is that the likelihood we obtain for the SMEFT parameters is Gaussian.
All in all, we provide simultaneous  constraints on 61 linear combinations of the dimension-6 Wilson coefficients.  
In this paper we quote the central values, the 68\% confidence level (CL) intervals,  while the correlation matrix is provided in the attached {\tt Mathematica} notebook\cite{magicnotebook}. 
That file also contains the full likelihood function in an electronic form, so that it can be more easily integrated into other analyses.  
 
The outline of the paper is the following. \sref{formalism} introduces the theoretical framework and the necessary notation. \sref{lee} presents the experimental input of our analysis. \sref{gf} contains the results of our fit, in the general case and in some interesting limits. Finally \sref{lhc} discusses the interplay with LHC searches, and \sref{concl} contains our conclusions.

%%%%%%%%%%%%%%%%%%%%%%
\section{Formalism and notation}
\label{sec:formalism}
\setcounter{equation}{0}

%---------------------------------------------------------------------
\subsection{SMEFT with dimension-6 operators}

Our  framework is that of  the baryon- and lepton-number conserving SMEFT~\cite{Leung:1984ni,Buchmuller:1985jz}.
The Lagrangian is organized as an expansion in $1/\Lambda^2$, where $\Lambda$ is interpreted as  the mass scale of new particles in the UV completion of the effective theory. 
We truncate the expansion at $\cO(\Lambda^{-2})$, which corresponds to retaining operators up to the canonical dimension $D$=6 and neglecting operators with  $D \geq 8$.  
The Lagrangian takes the form
\begin{equation}
\label{eq:SMEFT_l}
{\cal L} = {\cal L}_{\rm SM}+ \sum_{i}  \frac{c_{i}}{v^2} O_{i}^{D=6}, 
\end{equation}
where $ {\cal L}_{\rm SM}$ is the SM Lagrangian, $v=(\sqrt{2}G_F)^{-1/2}\simeq 246$~GeV,  each $O_i^{D=6}$ is a gauge-invariant operator of dimension $D$=6, and $c_i$ are the corresponding Wilson coefficients that are $\cO(\Lambda^{-2})$.
 $O_{i}^{D=6}$ span the complete space of dimension-6 operators, see Refs.~\cite{Grzadkowski:2010es,Contino:2013kra} for examples of such sets.  
 
In order to connect the SMEFT to observables it is convenient to rewrite \eref{SMEFT_l} using the mass eigenstates after electroweak symmetry breaking. 
Then the effects of dimension-6 operators show up as corrections to the SM couplings between fermion, gauge and Higgs fields, or as new interaction terms not present in the SM Lagrangian. 
The discussion and notation below follows closely that in Section~II.2.1 of Ref.~\cite{yr4}. 
We define the mass eigenstates such that all kinetic and mass terms are diagonal and canonically normalized. 
We also redefine couplings such that, at tree level, the relation between the usual SM input observables $G_F$, $\alpha$, $m_Z$ and the Lagrangian parameters $g_L$, $g_Y$, $v$ is the same as in the SM. 
See Ref.~\cite{yr4} for complete definition of conventions and the complete list of interaction terms with up to 4 fields.  
In the following we only highlight the parts of the mass eigenstate Lagrangian directly relevant for the analysis in this paper. 

One important effect from the point of view of precision measurements is the shift of the interaction strength of  the weak bosons. 
We parametrize the interactions between the electroweak gauge bosons and fermions as
\bea
\label{eq:SMEFT_dg}
{\cal L} & \supset  & 
  e A^\mu \sum_{f=u,d,e} Q_f (  \bar f_I \bar \sigma_\mu f_I  + f^c_I \sigma_\mu  \bar f^c_I)    \nnl
& + & {g_L \over \sqrt 2}\left [ W^{\mu+}  \bar \nu_I \bar \sigma_\mu (\delta_{IJ} +  [\delta g^{W e}_L]_{IJ}  ) e_J+ W^{\mu+}  \bar u_I \bar \sigma_\mu \left(V_{IJ} +  \left[\delta g^{W q}_L\right]_{IJ}  \right) d_J +\hc \right ] \nnl
& + & {g_L \over \sqrt 2}\left [ W^{\mu+}  u_I^c  \sigma_\mu \left[\delta g^{W q}_R\right]_{IJ} \bar d_J^c +\hc \right ] \nnl 
&+& \sqrt{g_L^2 + g_Y^2} Z^\mu    
\sum_{f=u,d,e,\nu}\bar f_I \bar \sigma_\mu \left (  (T_3^f -s_\theta^2 Q_f) \delta_{IJ} + \left[\delta g^{Zf}_L\right]_{IJ} \right ) f_J \nnl
&+& \sqrt{g_L^2 + g_Y^2} Z^\mu \sum_{f=u,d,e} f_I^c \sigma_\mu \left (   - s^2_\theta Q_f \delta_{IJ} + \left[\delta g^{Z f}_R \right]_{IJ}\right ) \bar f_J^c .
\eea
Here, $g_L$, $g_Y$ are the gauge couplings  of the $SU(2)_L \times U(1)_Y$ local symmetry, the electric coupling is $e = g_L g_Y/\sqrt{g_L^2 + g_Y^2}$, the sine of the weak mixing angle is $s_\theta = g_Y/\sqrt{g_L^2 + g_Y^2}$, and $I,J = 1,2,3$ are the generation indices. 
For the fermions we use the 2-component spinor formalism and we follow the conventions of Ref.~\cite{Dreiner:2008tw}, unless otherwise noted.\footnote{%
Compared to \cite{Dreiner:2008tw}, we use a different normalization of the antisymmetric product of the $\sigma$ matrices: 
$\sigma_{\mu\nu} = {i \over 2}(\sigma_\mu \bar \sigma_\nu - \sigma_\nu \bar \sigma_\mu)$,
 $\bar \sigma_{\mu\nu} = {i \over 2}(\bar \sigma_\mu  \sigma_\nu - \bar \sigma_\nu \sigma_\mu)$. }
The SM fermions $f_J$, $f^c_J$ are in the basis where the mass terms are diagonal, and then the CKM matrix $V$ appears in  the quark doublets as $q_I = (u_I, V_{IJ} d_J)$.  
The effects of dimension-6 operators are parameterized by the vertex corrections $\delta g$ that in  general can be flavor-violating.  
For flavor-diagonal interactions we will employ the shorter notation $[\delta g^{Vf}_{L/R}]_{JJ} \equiv \delta g^{V f_J}_{L/R}$.

The vertex corrections can be expressed as linear combinations of the Wilson coefficients $c_i$ in \eref{SMEFT_l}, see \aref{warsaw} for the map to the Warsaw basis. 
We find more transparent to recast the results of precision experiments as constraints on $\delta g$'s. 
This is completely equivalent, provided one takes into account that not all $\delta g$'s in \eref{SMEFT_dg} are independent.\footnote{%
More generally, it is often convenient to parametrize the space of dimension-6 operators using $\delta g$'s and other independent parameters  in the mass eigenstate Lagrangian that are in a 1-to-1 linear relation with the set of Wilson coefficients $c_i$ \cite{Gupta:2014rxa}. 
One example of such parametrization goes under the name of the Higgs basis and is defined in Ref.~\cite{yr4}.
}
Indeed, the mapping between the vertex corrections and the Wilson coefficients implies the relations 
$[\delta g^{Z\nu}_L]_{IJ} -  [\delta g^{Ze}_L]_{IJ}  = [\delta g^{W e}_L]_{IJ}$, 
and $ [\delta g^{W q}_L]_{IJ} = [\delta g^{Z u}_L]_{IK}V_{KJ} -  V_{IK}[\delta g^{Zd}_L]_{KJ}$. 

In this paper we focus on flavor-conserving observables that target flavor-diagonal Wilson coefficients. We will express the experimental constraints using the following set of independent flavor-diagonal vertex corrections: 
\beq
\label{eq:SMEFT_dgset}
 \delta g^{Ze_I}_L, \,  \delta g^{Ze_I}_R, \,  \delta g^{W e_I}_L, \, 
 \delta g^{Z u_I}_L, \,  \delta g^{Z u_I}_R,  \, \delta g^{Z d_I}_L, \,  \delta g^{Z d_I}_R,  \,  \delta g^{W q_I}_R . 
\eeq 
The vertex corrections correspond to 24 linear combinations of dimension-6 Wilson coefficients, 3 of which are complex (those entering $\delta g_{R}^{Wq}$). We consider only CP-conserving observables, thus the imaginary part enters at the quadratic level and is neglected.
To simplify the notation we will omit $\re$ in front of complex Wilson coefficients.

\begin{table}
\bc
\begin{tabular}{c|c}
\hline 
Chirality conserving ($I,J = 1,2,3$) &  Chirality violating ($I,J = 1,2,3$) \\ 
\hline 
\\
$ [O_{\ell q}]_{IIJJ} = (\bar \ell_I\bar \sigma_\mu \ell_I)  (\bar q_J \bar \sigma^\mu q_J)$
 & $ [O_{\ell e q u}]_{IIJJ}  =  (\bar \ell^j_I \bar e_I^c) \epsilon_{jk} (\bar q^k_J \bar u^c_J) $
 \\ 
$ [O^{(3)}_{\ell q}]_{IIJJ} = (\bar \ell_I \bar \sigma_\mu \sigma^i \ell_I)  (\bar q_J \bar \sigma^\mu \sigma^i q_J)$ 
& $ [O^{(3)}_{\ell e q u}]_{IIJJ}  =  
 (\bar \ell^j_I \bar \sigma_{\mu \nu} \bar e_I^c) \epsilon_{jk} (\bar q^k_J  \bar \sigma_{\mu \nu} \bar u^c_J) $
 \\   
$ [O_{\ell u}]_{IIJJ} = (\bar \ell_I\bar \sigma_\mu \ell_I)  (u^c_J \sigma^\mu \bar u^c_J)$ 
 & $ [O_{\ell e d q}]_{IIJJ}  =  (\bar \ell^j_I \bar e_I^c)  (d^c_J q^j_J) $ 
 \\
$ [O_{\ell d}]_{IIJJ} = (\bar \ell_I\bar \sigma_\mu \ell_I)  (d^c_J \sigma^\mu \bar d^c_J)$ &
  \\
$ [O_{e q}]_{IIJJ} = (e^c_I \sigma_\mu \bar e^c_I)  (\bar q_J \bar \sigma^\mu q_J)$ &
\\ 
$ [O_{e u}]_{IIJJ} = (e^c_I \sigma_\mu \bar e^c_I)  (u^c_J \sigma^\mu \bar u^c_J)$ &
\\ 
$ [O_{e d}]_{IIJJ} = (e^c_I \sigma_\mu \bar e^c_I)  (d^c_J \sigma^\mu \bar d^c_J)$ &
\\ 
 \end{tabular}
\ec 
\caption{Flavor-conserving {\bf 2-lepton-2-quark} operators in the SMEFT Lagrangian  of \eref{SMEFT_l}.
\label{tab:2l2q}
}
\end{table}
\begin{table}
\bc
\begin{tabular}{c|c}
\hline 
One flavor ($I=1,2,3$) & Two flavors ($I < J =1,2,3$) \\ 
\hline 
&
\\
$ [O_{\ell \ell}]_{IIII} = {1\over 2} (\bar \ell_I\bar \sigma_\mu \ell_I)  (\bar \ell_I \bar \sigma^\mu \ell_I)$
 & $ [O_{\ell \ell}]_{IIJJ}  =  (\bar \ell_I\bar \sigma_\mu \ell_I)  (\bar \ell_J \bar \sigma^\mu \ell_J) $ 
 \\ 
  & $[O_{\ell \ell}]_{IJJI} = (\bar \ell_I \bar \sigma_\mu \ell_J)  (\bar \ell_J \bar \sigma^\mu \ell_I)  $
 \\   
$ [O_{\ell e}]_{IIII} =  (\bar \ell_I\bar \sigma_\mu \ell_I)  (e_I^c  \sigma^\mu \bar e_I^c) $ &
$ [O_{\ell e}]_{IIJJ}  =  (\bar \ell_I\bar \sigma_\mu \ell_I)  (e_J^c  \sigma^\mu \bar e_J^c)$
 \\
 &  $[O_{\ell e}]_{JJII}  =  (\bar \ell_J \bar \sigma_\mu \ell_J)  (e_I^c  \sigma^\mu \bar e_I^c)$
  \\
 &  $[O_{\ell e}]_{IJJI}  =  (\bar \ell_I \bar \sigma_\mu \ell_J)  (e_J^c  \sigma^\mu \bar e_I^c)$
\\ 
 $ [O_{e e}]_{IIII} =   {1\over 2} (e_I^c  \sigma_\mu \bar e_I^c)   (e_I^c  \sigma^\mu \bar e_I^c) $ &
 $ [O_{e e}]_{IIJJ}  =   (e_I^c  \sigma_\mu \bar e_I^c)   (e_J^c  \sigma^\mu \bar e_J^c) $
 \end{tabular}
\ec 
\caption{Flavor-conserving {\bf 4-lepton operators} in the SMEFT Lagrangian of \eref{SMEFT_l}.
\label{tab:4l}
}
\end{table}  

In this paper we will also discuss  constraints on flavor-diagonal 4-fermion operators in the SMEFT Lagrangian of \eref{SMEFT_l}. 
We work with the same set of 4-fermion operators as in Ref.~\cite{Grzadkowski:2010es} and employ a similar notation.\footnote{%
One difference is that for operators with the $SU(2)_L$ singlet contraction of fermionic currents we omit the superscript~${}^{(1)}$. 
We also rename ${\cal Q}_{qe}\to {\cal Q}_{eq}$ so that the first (last) two flavor indices of all LLQQ operators correspond to the leptons (quarks).
}
The main focus is on the flavor-conserving 2-lepton-2-quark dimension-6 operators (LLQQ) summarized in \tref{2l2q}, and defined in the flavor basis where the up-quark Yukawa matrices are diagonal. 
Overall, there are $10 \times 3 \times 3 = 90$ such operators, of which  27 (the chirality-violating ones) are complex. In the latter case the corresponding Wilson coefficient is complex, and the Hermitian conjugate operator is included in \eref{SMEFT_l}. 
For the sake of combining our results with those of Ref.~\cite{Falkowski:2015krw},  
we also list in \tref{4l} the 27 flavor-conserving  4-lepton operators (LLLL), 3 of which are complex ($[O_{\ell e}]_{IJJI}$).  

All in all, our analysis eyes $147$ linear combinations of dimension-6 operators displayed in \eref{SMEFT_dgset}, \tref{2l2q}, and \tref{4l}. 
The observables discussed in this paper will not depend on all of them, and thus we will be able to constrain only  a limited number of the combinations.
In particular  the operators involving the 3rd generation fermions are currently, with a few exceptions,  poorly constrained  by experiment.  
Nevertheless, the constraints we derive are robust, in the sense that they do not involve any strong assumptions about the unconstrained operators, other than the validity of the SMEFT description at the weak scale. 
We assume that our results are not invalidated by $\cO\left({1 \over 16 \pi^2 \Lambda^2} \right )$ corrections, which arise at one loop in the SMEFT and inevitably introduce dependence of our observables on other $D$=6 Wilson coefficients. 
We will also treat $V$ as the unit matrix when it multiplies dimension-6 Wilson coefficients. 
This ignores all contributions to observables where the Wilson coefficients are  multiplied by an off-diagonal CKM element.\footnote{%
Such an approach is not completely satisfactory, since the Cabibbo angle is not small enough to always justify neglecting it. 
However, including the new physics contributions suppressed by the Cabibbo angle would require extending  our analysis to include  flavor-violating observables, which we leave for future publications. On the other hand, one naively expects the neglected operators  to be severely constrained by other observables where the CKM suppression is not present, which would justify our approximation.}

Last, we will also particularize our results to more restrictive scenarios, such as the so-called flavor-universal SMEFT, where dimension-6 operators respect the $U(3)^5$ global flavor symmetry acting in the generation space on the SM fermion fields $q$, $\ell$, $u^c$, $d^c$, $e^c$.

%---------------------------------------------------------------------
\subsection{Weak interactions below the weak scale}

Precision experiments with a characteristic momentum transfer $Q \ll m_Z$ can be conveniently described using the low-energy effective theory where the SM $W$ and $Z$ bosons are integrated out. 
In this framework, weak interactions between quark and leptons are mediated by a set of 4-fermion operators.  
Within the SM, these operators effectively appear due to the exchange of  $W$ and $Z$ bosons at tree level or in loops, and their coefficients can be calculated by the standard matching procedure. 
Once the SM is extended by dimension-6 operators, these coefficients may be modified, either due to modified propagators and couplings of $W$ and $Z$, or due to the presence of contact 4-fermion operators in the SMEFT Lagrangian. 

Below we define the low-energy operators that are relevant for the precision measurements we include in  our analysis. 
We follow the PDG notation~\cite{Olive:2016xmw} (Section~10), and we present the matching between the coefficients of the low-energy operators and  the parameters of the SMEFT.   

\subsubsection{Charged-current (CC) interactions: $qq'\ell\nu$}
\label{sec:CC}
The low-energy CC interactions of leptons with the 1st generation quarks are described by the effective 4-fermion operators:
\bea 
\label{eq:LE_nucc}
{\cal L}_{\rm eff} & \supset &
-\frac{2 \tilde V_{ud}}{v^2} \left [  
\left ( 1+  \eL^{d e_J} \right ) (\bar e_J  \bar \sigma_\mu \nu_J)(\bar u \bar \sigma^\mu d) 
+ \eps_R^{d e}   (\bar e_J  \bar \sigma_\mu \nu_J)(u^c \sigma^\mu \bar d^c) 
\right . \\ && \left . 
+ {\eps_S^{d e_J}  +\eps_P^{d e_J} \over 2} (e_J^c  \nu_J) (u^c d) 
+ {\eps_S^{d e_J}  - \eps_P^{d e_J} \over 2}  (e_J^c  \nu_J) (\bar u \bar d^c)  
%\right . \nnl && \left . 
+ \eps_T^{d e_J}  (e_J^c   \sigma_{\mu \nu}  \nu_J) (u^c \sigma_{\mu \nu} d) 
+ \hc  \right ]. \nonumber
\eea 
To make contact with low-energy flavor  observables, we defined the rescaled CKM matrix element $\tilde V_{ud}$~\cite{Gonzalez-Alonso:2016etj}. 
It is distinct from the \emph{actual} $V_{ud}$, i.e., the $11$ element of the unitary matrix $V$ that appears in the Lagrangian after rotating  quarks  to the mass eigenstate basis.  
The two are related by $V_{ud} = \tilde V_{ud} (1 + \delta V_{ud})$ where $\delta V_{ud}$ is chosen such as to impose  the relation $\eL^{d e} =   - \eps_R^{de}$ in \eref{LE_nucc}.\footnote{The bar in the $\eL^{d e_J}$ coefficient reminds the reader that this coefficient is not the usual $\epsilon_L^{d e_J}$ (see e.g. Ref.~\cite{Gonzalez-Alonso:2016etj}) where the shift of new physics effects into $\tilde V_{ud}$ is not carried out. These two are trivially related by $V_{ud}\,(1+\epsilon_L^{d e_J}) = \tilde V_{ud}\, (1+\eL^{d e_J})$.}

Let us note that in general $\tilde V_{ud}$ is also different from the phenomenological value obtained within the SM, which we will denote by $V_{ud}^{\rm{PDG}}$. Currently this value comes from superallowed nuclear beta decays~\cite{Hardy:2014qxa} that depend on the vector couplings via the combination $\eL^{d e} + \eps_R^{d e}$. By setting $\eL^{d e} =   - \eps_R^{d e}$, this nonstandard effect has been conveniently absorbed into the definition of $\tilde V_{ud}$. However, the relevant transitions also depend, each in  a different way, on the scalar coefficient $\epsilon_S^{de}$. Thus $\tilde V_{ud}$ and $V_{ud}^{\rm{PDG}}$ only coincide if $\epsilon_S^{de}$ vanishes, whereas in general it is not possible to redefine away all new physics contributions through $\tilde V_{ud}$. For this reason we treat  $\tilde V_{ud}$ as a free parameter that is fit together with the EFT Wilson coefficients~\cite{Gonzalez-Alonso:2016etj}. In principle the difference between $\tilde V_{ud}$ and $V_{ud}^{\rm{PDG}}$ must be taken into account every time the latter is used to calculate any given SM prediction. In practice, this effect will be negligible in most cases, given the strong constraints on $\eps_{S}^{de}$ from the same nuclear decay data, \emph{cf.} \eref{6Dleffe}.

At tree level, the low-energy parameters are related to the SMEFT parameters as
\bea
\delta V_{ud} & = &    - \delta g_L^{W q_1}  - \delta g_R^{W q_1} +  \delta g_L^{W \mu} - {1\over 2} [c_{\ell \ell}]_{1221}   +  [c^{(3)}_{lq}]_{1111}, \nnl
\label{eq:epsilon}
\epsilon_R^{d e}  =  - \eL^{d e} &= &  \delta g_R^{W q_1} , 
\nnl 
 \eL^{d \mu}   &= &  -  \delta g_R^{W q_1}  +  \delta g_L^{W \mu}  -  \delta g_L^{W e} + [c^{(3)}_{lq}]_{111 1} -  [c^{(3)}_{lq}]_{2211}, 
\nnl 
\epsilon_S^{d e_J} &=& - \frac{1}{2} \left ( [c_{lequ}]^*_{JJ11} +  [c_{ledq}]^{*}_{JJ11} \right ),
\nonumber\\
\epsilon_P^{d e_J} &=& -  \frac{1}{2} \left (   [c_{lequ}]^*_{JJ11} -  [c_{ledq}]^{*}_{JJ11} \right ),  
\nonumber\\
\epsilon_T^{d e_J} &=& - \frac{1}{2} [c^{(3)}_{lequ}]^*_{JJ11}~,
\eea 
As indicated earlier, at $\cO(\Lambda^{-2})$ we treat the CKM matrix as the unit matrix. 
In this limit, the effective parameters in  \eref{LE_nucc} depend only on flavor-diagonal vertex corrections and 4-fermion operators. 
See \aref{LEFFE} for more general expressions where non-diagonal elements of $V$ are retained. 
Note also that the rescaled CKM matrix is no longer unitary.  In particular we have $|\tilde V_{ud}|^2  + |V_{us}|^2   \approx 1 + \Delta_{\rm CKM}$, where 
\bea 
\label{eq:deltackm}
\Delta_{\rm CKM}  =  -2\delta V_{ud} =   2 \delta g_L^{W q_1}  + 2 \delta g_R^{W q_1} -  2 \delta g_L^{W \mu} +  [c_{\ell \ell}]_{1221} - 2 [c^{(3)}_{lq}]_{1111}.
\eea 
Although the extraction of the $V_{us}$ element is also affected by dimension-6 operators, their contribution to this unitarity test is suppressed by $V_{us}$ and therefore it can be neglected in our approximation ($V \approx 1$ at order $\Lambda^{-2}$). See \eref{deltackm1} for the complete expression.

%---------------------------------------------------------------------------------------------------
\vspace{0.1cm}
\subsubsection{Neutral-current (NC) neutrino interactions: $qq\nu\nu$}

The low-energy NC neutrino interactions with light quarks are described by the effective 4-fermion operators:
\bea 
\label{eq:LE_nunc} 
{\cal L}_{\rm eff} & \supset & -  \frac{2}{v^2}(\bar \nu_J \bar \sigma^\mu \nu_J) \left ( 
g_{LL }^{\nu_J q}  \bar q\bar \sigma_\mu q + g_{LR }^{\nu_J q } q^c \sigma_\mu \bar q^c \right ). 
\eea
At tree level, the low-energy parameters are related to the SMEFT parameters as 
\bea 
\label{eq:LE_nunc2} 
g_{LL}^{\nu_J u} &=&{1 \over 2 } -  {2 s_\theta^2 \over 3 }  +  \delta g_L^{Zu} +
\left ( 1 - {4 s_\theta^2 \over 3 }  \right ) \delta g_L^{Z\nu_J} -\frac{1}{2}([c_{lq}]_{JJ11}+[c^{(3)}_{lq}]_{JJ11}) ,
\nnl 
g_{LR }^{\nu_J u} &=& - {2 s_\theta^2 \over 3}  
 + \delta g_R^{Zu}  - {4 s_\theta^2 \over 3} \delta g_L^{Z\nu_J}-\frac{1}{2}[c_{lu}]_{JJ11} ,
\nnl 
g_{LL }^{\nu_J d} &=& - {1 \over 2 } + {s_\theta^2 \over 3 }  +  \delta g_L^{Zd}  - 
\left (  1 -   {2 s_\theta^2  \over 3 }  \right ) \delta g_L^{Z\nu_J} 
-\frac{1}{2}([c_{lq}]_{JJ11}-[c^{(3)}_{lq}]_{JJ11}) ,
\nnl 
g_{LR }^{\nu_J d} &=& {s_\theta^2 \over 3} + \delta g_R^{Zd} +{2 s_\theta^2 \over 3 }  \delta g_L^{Z\nu_J} 
-\frac{1}{2}[c_{ld}]_{JJ11}.  
\end{eqnarray}
The experiments probing these couplings usually normalize the NC cross section using its CC counterpart. Thus,  it is convenient to define the following combinations of effective couplings:  
\beq
\label{eq:LE_nunc3} 
(g_{L/R}^{\nu_J})^2  \equiv    {(g_{LL/LR}^{\nu_J u})^2 + (g_{LL/LR}^{\nu_J d})^2 \over 
 \left ( 1+ \eL^{d e_J} \right )^2 }, 
\qquad 
\theta_{L/R}^{\nu_J}   \equiv   \arctan \left( g_{LL/LR}^{\nu_J u} \over g_{LL/LR}^{\nu_J d} \right ),
\eeq
where we took into account that SMEFT dimension-6 operators modify in general both NC and CC processes. 
Let us notice that additional (linear) effects in the normalizing CC process due to $\eps_{R}^{d e}$ and $\eps_{S,P,T}^{d e_J}$ can be neglected because they are suppressed by the ratio $m_u m_d / E^2$ and $m_{e_J} / E$ respectively. The effect due to the possible difference between  $\tilde V_{ud}$ and $V_{ud}^{\rm PDG}$ can also be safely neglected here, given the limited precision of the neutrino scattering experiments included in our fit. Last, the same holds for the $\delta V_{ud}$ contribution that appears if the unitarity of the CKM matrix is used in the SM determination.

\vspace{0.1cm}
\subsubsection{Neutral-current charged-lepton interactions: $qq\ell\ell$}

We parametrize\footnote{%
For the parity-violating electron couplings, another frequently used   notation is $g_{AV}^{e q} \equiv C_{1q}$, $g_{VA}^{e q} \equiv C_{2q}$. }
the 4-fermion operators with 2 charged leptons and 2 light quarks as 
\bea
\cL &\supset& 
{1 \over 2 v^2} \left [ 
g^{e_J q}_{AV} (\bar e_J \gamma_\mu\gamma_5   e_J ) (\bar q \gamma_\mu q) 
+ g^{e_J q}_{VA} (\bar e_J \gamma_\mu  e_J ) (\bar q \gamma_\mu \gamma_5  q)  
\right]\nnl
&&\!\!\!\!\!+\,{1 \over 2 v^2} \left [ 
g^{e_J q}_{VV} (\bar e_J \gamma_\mu  e_J ) (\bar q \gamma_\mu   q)  
+ g^{e_J q}_{AA} (\bar e_J \gamma_\mu  \gamma_5 e_J ) (\bar q \gamma_\mu \gamma_5  q)  
\right ], 
\eea 
where we momentarily switch to the Dirac notation with 
$\gamma_5 \psi_L = - \psi_L $,  $\gamma_5 \psi_R = +\psi_R$. 
At tree level, the parameters $g^{e_i q}_{XY}$ are related to the SMEFT parameters as 
\bea
\label{eq:LE_ge}
g^{e_J u}_{AV}   & =&  -{1 \over 2} + {4 \over 3} s_\theta^2  
-  \left ( \delta g^{Zu}_L +   \delta g^{Zu}_R  \right ) 
+ \frac{3 - 8 s_\theta^2}{3} \left ( \delta g^{Ze_J}_L -  \delta g^{Ze_J}_R \right )  
+ {1 \over 2} \left[ c_{lq}^{(3)}  -   c_{lq}  -   c_{lu} +   c_{eq}  +  c_{eu} \right]_{JJ11} , 
\nnl 
g^{e_J d}_{AV}  & =& {1 \over 2}  -  {2 \over 3} s_\theta^2 
 -  \left ( \delta g^{Zd}_L +   \delta g^{Zd}_R  \right ) 
- \frac{3 - 4 s_\theta^2}{3} \left ( \delta g^{Ze_J}_L -  \delta g^{Ze_J}_R \right )  
+{1 \over 2} \left[  - c_{lq}^{(3)}  -   c_{lq}  -   c_{ld} +  c_{eq}  +  c_{ed} \right]_{JJ11},  
\nnl 
g^{e_J u}_{VA}   & =& -{1 \over 2} + 2 s_\theta^2  
 - \left ( 1 - 4 s_\theta^2 \right ) \left ( \delta g^{Zu}_L -  \delta g^{Zu}_R \right )  
+ \left (\delta g^{Ze_J}_L +  \delta g^{Z e_J}_R \right )
+{1 \over 2} \left[  c_{lq}^{(3)}  -  c_{lq} +   c_{lu} -  c_{eq}  +  c_{eu} \right]_{JJ11}, 
\nnl 
 g^{e_J d}_{VA}  & =& {1 \over 2}  - 2 s_\theta^2   - \left ( 1 - 4 s_\theta^2 \right ) \left ( \delta g^{Zd}_L -  \delta g^{Zd}_R \right )  
- \left (\delta g^{Ze_J}_L +  \delta g^{Ze_J}_R \right )
+{1 \over 2} \left[ -c_{lq}^{(3)}  -  c_{lq}  +  c_{ld} -  c_{eq}  +  c_{ed} \right]_{JJ11},
\nnl 
g^{e_J u}_{AA}   & =& 
{1 \over 2 } +  \delta g^{Zu}_L -  \delta g^{Zu}_R  -   \delta g^{Ze_J}_L +  \delta g^{Z e_J}_R 
+{1 \over 2} \left[  - c_{lq}^{(3)}  + c_{lq} -  c_{lu} -  c_{eq}  +  c_{eu} \right]_{JJ11}, 
\nnl 
g^{e_J d}_{AA}   & =& 
-{1 \over 2 } +   \delta g^{Zd}_L -  \delta g^{Zd}_R  +  \delta g^{Ze_J}_L -  \delta g^{Z e_J}_R 
+{1 \over 2} \left[  c_{lq}^{(3)}  + c_{lq} -  c_{ld} -  c_{eq}  +  c_{ed} \right]_{JJ11}.
\eea 
We do not display  the expressions for $g^{e_i q}_{VV}$ here because they will not be needed in the following.  
%The SM values (including once again known loop corrections) are~\cite{Olive:2016xmw}: 
%\beq
%g^{e u, \rm SM}_{AV} =  -0.1887, 
%\qquad 
%g^{e d, \rm SM}_{AV} =  0.3419,
%\qquad 
%g^{e u, \rm SM}_{VA} =  -0.0351, 
%\qquad 
%g^{e d, \rm SM}_{VA} =  0.0247, 
%\eeq 
%and $g^{e u, \rm SM}_{AA}  = - g^{e d, \rm SM}_{AA} = 1/2$ {at tree level}.
%%%%%%%%%%%%%%%%%%%%%%

\vspace{0.1cm}
\subsubsection{Four-lepton interactions: $\ell\ell\ell\ell$ and $\ell\ell\nu\nu$}
Although the main focus of this work are the LLQQ operators, in this section we provide a few expressions concerning 4-lepton operators that will be needed in our subsequent phenomenological analysis. First, we parametrize the $\nu$-$e$ interaction in the effective theory below the weak scale as:
\beq
\cL \supset  -  {1 \over  v^2}  (\bar \nu_J \bar \sigma_\mu \nu_J) \left [
\left ( g_{LV}^{\nu_J e_I} + g_{LA}^{\nu_J e_I}  \right ) (\bar e_I \bar \sigma_\mu e_I)
+ \left ( g_{LV}^{\nu_J e_I} -  g_{LA}^{\nu_J e_I}  \right )   (e_I^c \sigma_\mu \bar e_I^c)   \right ]. 
\eeq 
Matching to the SMEFT one finds   
\bea
g_{LV}^{\nu_J e_I}  & = &
 \delta_{IJ} -\frac{1}{2} + 2s_\theta^2 
 +\delta_{IJ} \left( 2  \delta g^{We_I}_L - \delta g^{We}_L - \delta g^{W\mu}_L + \frac{1}{2} [c_{\ell \ell}]_{1221}  \right)
 \nnl
 && -  \left ( 1 - 4 s_\theta^2 \right ) \delta g^{Z \nu_J}_L+  \delta g^{Ze_I}_L +  \delta g^{Ze_I}_R  - {1 \over 2 } \left ( x_{IJ} + [c_{\ell e}]_{JJII} \right ), 
  \nnl 
  g_{LA}^{\nu_J e_I} & =& 
   \delta_{IJ} - \frac{1}{2} 
 +\delta_{IJ} \left( 2  \delta g^{We_I}_L - \delta g^{We}_L - \delta g^{W\mu}_L + \frac{1}{2} [c_{\ell \ell}]_{1221}  \right)
  \nnl
 && -   \delta g^{Z \nu_J}_L +  \delta g^{Ze_I}_L -  \delta g^{Ze_I}_R
 - {1 \over 2 } \left ( x_{IJ} -  [c_{\ell e}]_{JJII} \right ),  
\eea 
where $x_{IJ}= [c_{\ell \ell}]_{IIJJ}$ if $I\le J$ or $x_{IJ}= [c_{\ell \ell}]_{JJII}$ otherwise. 
%$g_{LV}^{\nu e, \rm SM} =  -0.0396$,  $g_{LA}^{\nu e, \rm SM} = -0.5064$~\cite{Olive:2016xmw}.   

Last, we parameterize the parity-violating self-interaction of electrons in the effective theory below the weak scale as
\beq
\cL \supset   {1 \over 2 v^2} g_{AV}^{ee} \left [
- (\bar e \bar \sigma_\mu e)  (\bar e \bar \sigma_\mu e)
+  (e^c \sigma_\mu \bar e^c)  (e^c \sigma_\mu \bar e^c) \right ]~,
\eeq  
with the following SMEFT expression
\beq 
g_{AV}^{ee}  =  
\frac{1}{2} -2 s_\theta^2 
 - 2  \left ( 1 - 2 s_\theta^2 \right ) \delta g^{Ze}_L - 4 s_\theta^2 \delta g^{Ze}_R
 - {1 \over 2} [c_{\ell \ell}]_{1111} +  {1 \over 2}  [c_{e e}]_{1111} ~.
\eeq 
%where  $g_{AV}^{e e, \rm SM} =  0.0225$~\cite{Olive:2016xmw}.

\subsection{Renormalization and scale running of the Wilson coefficients}
\label{sec:rge}

In general the Wilson coefficients display renormalization-scale dependence that is to be canceled in the observables by the opposite dependence in the quantum corrections to the matrix elements. Let us first discuss the QCD running, which can have a numerically significant impact due to the magnitude of the strong coupling constant $\alpha_s$. This effect is further enhanced by the large separation of scales of the experiments discussed in this work (from low-energy precision measurements to LHC collisions). Among the coefficients involved in our analysis, only the three chirality-violating ones, $c_{lequ}, c_{ledq}, c^{(3)}_{lequ}$ (i.e. $\epsilon_{S,P,T}^{d\ell}$ in the low-energy EFT), present a non-zero 1-loop QCD anomalous dimension, namely~\cite{Eichten:1989zv}
\bea
\label{eq:qcdrunning}
\frac{d\,\vec x(\mu)}{d\log\mu}=  \frac{\alpha_s(\mu)}{2\pi}
\left(
\begin{array}{ccc}
 -4 & 0 & 0 \\
 0 & -4 & 0 \\
 0 & 0 & 4/3 \\
\end{array}
\right  )\,\vec x(\mu),
\eea
where $\vec x$ refers to the SMEFT coefficients $\vec c=(c_{ledq},\,c_{lequ},\,c^{(3)}_{lequ})$ if the scale $\mu$ is above the weak scale or to the low-energy EFT coefficients $\vec\epsilon=(\eS^{d\ell},\,\eP^{d\ell},\,\eT^{d\ell})$ below it.
We find that higher-loop QCD corrections to the running are numerically significant, and we include them in our  analysis.\footnote{
We use the 3-loop QCD anomalous dimension~\cite{Gracey:2000am}, and we include the threshold corrections at $m_b$ and $m_t$ extracted from  Refs.~\cite{Chetyrkin:1997un} and \cite{Misiak:2010sk} for scalar and tensor operators respectively. See Ref.~\cite{Gonzalez-Alonso:2017iyc} for further details.
}

On the other hand we neglect in this work the electromagnetic/weak running of the SMEFT Wilson coefficients, which is expected to have a much smaller numerical importance simply due to the smallness of the corresponding coupling constants. There is however one exception to this, namely the chirality-violating operators discussed above, for two reasons: (i) contrary to the QCD running, the QED/weak running involves mixing between these operators; (ii) pion decay makes possible to set bounds of order $10^{-7}$ on the pseudoscalar coupling $\eP^{d\ell}(\mu_{\mbox{low}})$, which can give important bounds on scalar and tensor via mixing despite the smallness of $\alpha_{em}$. In order to take into account this effect, \eref{qcdrunning} has to be replaced by
\bea
\label{eq:qedrunning}
\frac{d\,\vec x(\mu)}{d\log\mu}=   \left( \frac{\alpha_{em}(\mu)}{2\pi}\gamma_x + \frac{\alpha_s(\mu)}{2\pi}\gamma_s \right) \,\vec x(\mu)~,
\eea
where we will use the 1-loop QED (electroweak) anomalous dimension, $\gamma_x=\gamma_{em(w)}$, to evolve the coefficients $\vec \epsilon$ ($\vec c$) below (above) the weak scale~\cite{Celis:2017hod,Aebischer:2017gaw,Gonzalez-Alonso:2017iyc,Alonso:2013hga}:
\bea
\gamma_{\rm em}=
\left(
\begin{array}{ccc}
 \frac{2}{3} & 0 & 4 \\
 0 &  \frac{2}{3} & 4 \\
 \frac{1}{24} &  \frac{1}{24} &  -\frac{20}{9} \\
\end{array}
\right)~,~~~~~
\gamma_{\rm w}=
\left(
\begin{array}{ccc}
 -\frac{4}{3c_\theta^2}		&	0					&		0		\\
 0						& -\frac{11}{6c_\theta^2}		&	\frac{15}{c_\theta^2}+\frac{9}{s_\theta^2}  \\
 0						& \frac{5}{16c_\theta^2} + \frac{3}{16s_\theta^2} & \frac{1}{9c_\theta^2} - \frac{3}{2s_\theta^2} \\
\end{array}
\right)
%=\left(
%\begin{array}{ccc}
% -1.73 & 0 & 0 \\
% 0 & -2.38 & 58.4 \\
% 0 & 1.22 & -6.34 \\
%\end{array}
%\right)
~,
\eea
where we neglect terms suppressed by Yukawa couplings~\cite{Jenkins:2013wua,Alonso:2013hga}.
Integrating numerically the coupled differential renormalization group equations we find
\bea
\left(
\begin{array}{c}
\epsilon_S^{d\ell}\\
\epsilon_P^{d\ell} \\
\epsilon_T^{d\ell} \\
\end{array}
\right)_{\mbox{($\mu=m_Z$)}}
\!\!\!\!&=&\!\!\!
\left(
\begin{array}{ccc}
 0.58 & 1.42\times 10^{-6} & 0.017 \\
 1.42\times 10^{-6} & 0.58 & 0.017 \\
 1.53\times 10^{-4} & 1.53\times 10^{-4} & 1.21 \\
\end{array}
\right) 
\left(
\begin{array}{c}
\epsilon_S^{d\ell}\\
\epsilon_P^{d\ell} \\
\epsilon_T^{d\ell} \\
\end{array}
\right)_{\mbox{($\mu=2$~GeV)}}\!\!,
\label{eq:RGEepsilon}\\
\left(
\begin{array}{c}
c_{ledq} \\
c_{lequ}\\
c^{(3)}_{lequ} \\
\end{array}
\right)_{\mbox{($\mu=1$ TeV)}}
\!\!\!\!&=&\!\!\!
\left(
\begin{array}{ccc}
 0.84 & 0 & 0 \\
 0 & 0.84 & 0.16 \\
 0 & 3.3 \times 10^{-3} & 1.04 \\
\end{array}
\right) 
\left(
\begin{array}{c}
c_{ledq}\\
c_{lequ} \\
c^{(3)}_{lequ} \\
\end{array}
\right)_{\mbox{($\mu=m_Z$)}}~. 
\label{eq:RGEsmeft}
\eea
These results use the QCD beta function and anomalous dimensions up to 3 loops, and we included the bottom and top quark thresholds effects, see Ref.~\cite{Gonzalez-Alonso:2017iyc} for details.  
The diagonal entries would change by $\sim 12\%$ if just 1-loop QCD running were included, 
while two-loop results differ by only $\sim 1.5\%$. 
In our subsequent analysis we will use the numerical results in \eref{RGEepsilon}  and \eref{RGEsmeft}. 

\section{Low-energy experiments}
\label{sec:lee}
\setcounter{equation}{0} 

%-------------------------------------------------------------
\subsection{Neutrino scattering}

Neutrino scattering experiments measure the ratio of neutral- and charged-current neutrino or anti-neutrino scattering cross sections on nuclei: 
\beq
R_{\nu_i} =  {\sigma(\nu_i N \to \nu X) \over  \sigma(\nu_i N \to \ell_i^-  X) }, 
\qquad 
R_{\bar \nu_i} =  {\sigma(\bar \nu_i N \to \bar \nu X) \over  \sigma(\bar \nu_i N \to \ell_i^+  X) }.
\eeq 
At leading order and for isoscalar nucleus targets (equal number of protons and neutrons) one has the so-called Llewellyn-Smith relations~\cite{LlewellynSmith:1983tzz}:
\beq
\label{eq:EXP_rnu}
R_{\nu_i} = (g_{L}^{\nu_i})^2 + r  (g_{R}^{\nu_i})^2, 
\qquad
R_{\bar \nu_i}  =  (g_{L}^{\nu_i})^2 + r^{-1}  (g_{R}^{\nu_i})^2  , 
\eeq
where $r$ is the ratio of $\nu$ to $\bar \nu$ charged-current cross sections on $N$ that can be measured separately, and the effective couplings $g_{L/R}^{\nu_i}$ are defined in \eref{LE_nunc3}. 
In some experiments the  beam is a mixture of neutrinos and anti-neutrinos, and the following ratio is measured 
\beq
%\label{eq:Rcharmnue}
R_{\nu_i \bar \nu_i}= {\sigma(\nu_i N \to \nu X) + \sigma(\bar \nu_i N \to \bar \nu X) \over 
\sigma(\nu_i N \to \ell_i^-  X) + \sigma(\bar \nu_i N \to \ell_i^+ X)} 
=   (g_{L}^{\nu_i})^2 +  (g_{R}^{\nu_i})^2 .
\eeq 

\textbf{$\nu_e$ data.-} The CHARM experiment \cite{Dorenbosch:1986tb} made a measurement of electron-neutrino scattering cross sections:  
\beq
%\label{eq:Rcharmnue}
R_{\nu_e \bar \nu_e}= 0.406^{+0.145}_{-0.135},
\eeq
where the uncertainties quoted here and everywhere else in this work are 1-sigma (68\%C.L.) errors. 
To avoid dealing with asymmetric errors we approximate it as $R_{\nu_e \bar \nu_e}= 0.41 \pm 0.14$, and we estimate  the SM expectation as  $R_{\nu_e \bar \nu_e}^{\rm SM} =0.33$.
To our knowledge, this weakly constraining measurement  is currently the best probe of the electron-neutrino neutral-current interactions.

\begin{table}
\begin{center}
\begin{tabular}{|l|c|c|c|c|}
  \hline
  Experiment & Observable & Experimental value & SM value & Ref.  \\
  \hline
  \multirow{2}{*}{CHARM ($r=0.456$)}
   & $R_{\nu_{\mu}}$ & $0.3093 \pm 0.0031$ & 0.3156 & \cite{Allaby:1987vr} \\
   & $R_{\bar \nu_\mu}$ & $0.390 \pm 0.014$ & 0.370  & \cite{Allaby:1987vr} \\\hline
  \multirow{2}{*}{CDHS ($r=0.393$) }
  & $R_{\nu_\mu}$ & $0.3072\pm 0.0033$ & 0.3091  & \cite{Blondel:1989ev}  \\
    & $R_{\bar \nu_\mu}$   & $ 0.382  \pm  0.016$ &  0.380  & \cite{Blondel:1989ev} \\ \hline
  CCFR & $\kappa$ & $0.5820\pm 0.0041$ & 0.5830 & \cite{McFarland:1997wx} \\
 % NuTeV & $g_L^2$ & $0.30005 \pm 0.00137$ & 0.3039\\
 % NuTeV & $g_R^2$ & $0.03076 \pm 0.00110$ & 0.0300\\
  \hline
\end{tabular}
\end{center}
\caption{The results of muon-neutrino scattering experiments most relevant for constraining dimension-6 operators in the SMEFT. The SM values of $R_{\nu_\mu}$ and $\kappa$ include subleading corrections~\cite{Erler:2013xha}, whereas those of $R_{\bar \nu_\mu}$ are the tree-level values, which should be sufficient taking into account the larger experimental errors.}
\label{tab:Rnumu}
\end{table}
 
\textbf{$\nu_\mu$ data.-} For the muon-neutrino scattering the experimental data are much more abundant and precise. 
We summarize the relevant results in \tref{Rnumu}. 
The observable $\kappa$ measured in CCFR probes the following combinations of couplings~\cite{McFarland:1997wx}:
\beq 
\kappa= 1.7897(g_L^{\nu_\mu})^2+1.1479(g_R^{\nu_\mu})^2 - {{0.0916  \left [ (g_{LL}^{\nu_\mu u})^2 - (g_{LL}^{\nu_\mu d})^2 \right ] + 0.0782  \left [ (g_{LR}^{\nu_\mu u})^2 -(g_{LR}^{\nu_\mu d})^2 \right ]} \over  {(1+ \bar \epsilon_{L}^{d \mu})^2}}~. 
\eeq  
The additional small dependence on the difference of the up and down effective couplings appears when one  takes into account that the target (in this case iron) is not exactly isoscalar.
For the reasons explained in Ref.~\cite{Olive:2016xmw}, in our fits we do not take into account the results of the NuTeV experiment. 

The observables in \tref{Rnumu} constrain 3 independent combinations of the SMEFT coefficients. Rather then combining these results ourselves, we use the PDG combination~\cite{Olive:2016xmw} that also uses additional experimental input~\cite{ErlerPrivateCommunication} from neutrino induced coherent neutral pion production from nuclei \cite{Kullenberg:2009pu,Grabosch:1985mt} and elastic neutrino-proton scattering~\cite{Horstkotte:1981ne,Ahrens:1986xe}. Although their precision is quite limited, their inclusion allows one to constrain the 4 muon-neutrino effective couplings to quarks~\cite{Erler:2013xha}. The results of the latest PDG fit are~\cite{Olive:2016xmw}:
\bea
(g_L^{\nu_\mu})^2 & = &  0.3005 \pm 0.0028,
\qquad 
(g_R^{\nu_\mu})^2  =    0.0329 \pm 0.0030, 
\nnl 
\theta_L^{\nu_\mu}  & = & 2.50  \pm  0.035, 
\qquad  \qquad 
\theta_R^{\nu_\mu}    =   4.56^{+0.42}_{-0.27}. 
\eea 
The correlations are quoted to be small in Ref.~\cite{Olive:2016xmw} and  in the following we neglect them. 
We symmetrize the uncertainty on $\theta_R$ taking  the larger of the errors, so as to avoid dealing with asymmetric errors. 
The corresponding SM predictions are given in \tref{inputsummary}. 
To evaluate their dimension-6 EFT corrections in \eref{LE_nunc2} we will use  $s_\theta^2 = 0.23865$, which is the central value in the $\ov{MS}$ scheme at low energies~\cite{Olive:2016xmw}. 
We neglect  the error of the SM predictions when it is much smaller than the experimental uncertainties; otherwise we combine it in quadrature.

We note that LLQQ (and 4-lepton) operators can also be probed via matter effects in neutrino oscillations, see e.g. \cite{Gonzalez-Garcia:2013usa,Coloma:2017egw}.
However, the resulting constraints are not available in the model-independent form where all 4-fermion  operators can be present simultaneously.
Moreover, neutrino oscillations probe linear combinations of lepton-flavor-diagonal operators and of the off-diagonal ones (which we marginalize over). 
For these reasons, we do not include the oscillation constraints in this paper. 

%    

%---------------------------------------------------------------------------------------------------------
\subsection{Parity violation in atoms and in scattering}

Atomic parity violation (APV) and parity-violating electron scattering experiments access the parity-violating effective couplings of electrons to quarks $g_{AV}^{e q}$ and $g_{VA}^{e q}$. In particular, APV and elastic scattering on a target with $Z$ protons and $N$ neutrons probe its so-called weak charge $Q_W$ that is given by
\beq
Q_W(Z,N) = - 2 \left ( (2Z+N) g^{e u}_{AV}  + (Z + 2N) g^{e d}_{AV}  \right ) ~,
\eeq  
up to small radiative corrections~\cite{Erler:2013xha,Olive:2016xmw}. The most precise determination is performed in $^{133}$Cs, where $Q_W(55,133-55)\approx -376 g^{e u}_{AV}  - 422 g^{e d}_{AV} $. Taking into account recent re-analyses~\cite{Dzuba:2012kx} of the measured parity-violating transitions in cesium atoms~\cite{Wood:1997zq}, the latest edition of the PDG Review~\cite{Olive:2016xmw} quotes 
\beq
Q_W^{\rm Cs} = -72.62 \pm 0.43,   
\eeq  
where the SM prediction is $Q_{W, \rm SM}^{\rm Cs}= -73.25 \pm 0.02$~\cite{Olive:2016xmw}. Other APV measurements, e.g. with thallium atoms, probe slightly different combinations of the $g^{e q}_{AV}$ couplings, although with larger errors.

Instead, a very different linear combination of $g^{e u}_{AV}$ and $g^{e d}_{AV}$ is precisely probed by measurements of the weak charge of the proton, $Q_W^{\rm p} = Q_W(1,0)$, in scattering experiments with low-energy polarized electrons.   
The QWEAK experiment \cite{Androic:2013rhu} finds
\beq
Q_W^{\rm p} = 0.064 \pm 0.012, 
\eeq 
where the SM prediction is $Q_{W,\rm SM}^{\rm p} =  0.0708 \pm 0.0003$~\cite{Olive:2016xmw}. 

In order to access  the effective couplings  $g_{VA}^{eq}$ one needs to resort to deep-inelastic scattering of polarized electrons. 
Currently, the most precise of these is the PVDIS  experiment  \cite{Wang:2014bba} that studies  electron scattering on deuterium targets. 
The experiment is sensitive to the following two linear combinations of effective couplings \cite{Wang:2014bba}:
 \bea
A_1^{\rm PVDIS}  & = &
  1.156 \times 10^{-4}\left ( 
  2  g^{e u}_{AV} -   g^{e d}_{AV} +     0.348 (2  g^{e u}_{VA} -   g^{e d}_{VA})  \right ) 
\nnl 
A_2^{\rm PVDIS}    & = &
  2.022 \times 10^{-4}  \left (  2  g^{e u}_{AV} -   g^{e d}_{AV}  +      0.594 (2  g^{e u}_{VA} -   g^{e d}_{VA}) \right )~.  
\eea 
The  measured values are \cite{Wang:2014bba}
\beq
A_1^{\rm PVDIS}  = (- 91.1 \pm 4.3) \times 10^{-6}, \qquad 
A_2^{\rm PVDIS}   = (- 160.8 \pm 7.1) \times 10^{-6},  
\eeq 
where the SM predictions are $A_{1,SM}^{\rm PVDIS} =  -(87.7\pm0.7) \times 10^{-6}$,  $A_{2,SM}^{\rm PVDIS} =  -(158.9\pm1.0) \times 10^{-6}$~\cite{Wang:2014bba}. 

The PDG combines the results of APV, QWEAK, and PVDIS experiments into correlated constraints on 3 linear combinations of $g_{VA}^{eq}$ and $g_{AV}^{eq}$ \cite{Olive:2016xmw}: 
\beq
\label{eq:PVES_pdg}
\bvec g^{eu}_{AV} + 2 g^{ed}_{AV}   \\ 2 g^{eu}_{AV} -  g^{ed}_{AV}     \\  2 g^{eu}_{VA} -  g^{ed}_{VA}   \evec 
= \bvec 0.489 \pm 0.005 \\  -0.708 \pm 0.016  \\ -0.144 \pm 0.068 \evec, 
\qquad 
\rho = \left (\ba{ccc} & -0.94  &  0.42 \\  & & -0.45 \\ \phantom{0} \ea \right ) .
\eeq
To disentangle  $g^{eu}_{VA}$ and   $g^{ed}_{VA}$ one needs more input from earlier (less precise) measurements of parity-violating scattering.
We include two results provided by the SAMPLE  collaboration \cite{Beise:2004py}:  
\beq
\label{eq:sample}
g^{eu}_{VA} - g^{ed}_{VA} = - 0.042 \pm 0.057, 
%0.040 \pm 0.035 \pm 0.02, 
\qquad 
g^{eu}_{VA} - g^{ed}_{VA} =  -0.12 \pm 0.074,
% 0.05 \pm 0.05 \pm 0.02 \pm 0.01.   
\eeq 
from the scattering of polarized electrons on deuterons in the quasi-elastic kinematic regime at two different values of the beam energy. 
Combining the likelihood obtained from \eref{PVES_pdg} with the SAMPLE results we find the following constraints: 
\beq
\bvec \delta  g^{eu}_{AV}   \\ \delta  g^{ed}_{AV}   \\  \delta g^{eu}_{VA}   \\  \delta  g^{ed}_{VA}   \\  \evec 
= \bvec 0.0033 \pm 0.0054 \\  -0.0047 \pm 0.0051 \\ -0.041 \pm   0.081 \\ -0.032 \pm 0.11 \evec, 
\qquad 
\rho = \left (\ba{cccc} & -0.98 & -0.37  &  -0.27 \\  &  & 0.37 & 0.27  \\ & &&  0.94 \\ \phantom{0} \ea \right ). 
\eeq
Here $\delta  g^{eq}_{XY}$  are shifts of the effective couplings away from the SM values, 
whose dependence on the dimension-6 Wilson coefficients can be read off from \eref{LE_ge}. 

There are also results concerning effective muon couplings to quarks. 
A CERN SPS experiment~\cite{Argento:1982tq} measured a DIS asymmetry using polarized muon and anti-muon scattering on an isoscalar carbon target. 
The results can be recast as the measurement of the observable $b_{\rm SPS}$ defined as 
\beq
b_{\rm SPS} = {3 \over 5 e^2 v^2} \left ( 
g^{\mu d}_{AA} - 2 g^{\mu u}_{AA} + \lambda(  g^{\mu d}_{VA} - 2 g^{\mu u}_{VA} ) \right ),  
\eeq 
where $\lambda$ is the muon beam polarization fraction. Two measurements of $b_{\rm SPS}$ at different beam energies and polarization fractions were carried out~\cite{Argento:1982tq}: 
\bea
b_{\rm SPS} & =&  -\left (1.47 \pm 0.42 \right ) \times 10^{-4} \, \mbox{GeV}^{-2} \ {\rm for} \ \lambda = 0.81\  \Rightarrow \ b_{\rm SPS}^{\rm SM} = -1.56 \times 10^{-4} \, \mbox{GeV}^{-2}~, 
\nnl 
b_{\rm SPS} & =&  -\left (1.74 \pm 0.81 \right ) \times 10^{-4}  \, \mbox{GeV}^{-2} \ {\rm for} \ \lambda = 0.66 \ \Rightarrow \ b_{\rm SPS}^{\rm SM} = -1.57 \times 10^{-4} \, \mbox{GeV}^{-2}~. 
\eea

%//////////////////////////////////////////////////////////////////////////////////////////////////
\subsection{Low-energy flavor}

The partonic process $d_j\to u_i \ell \bar{\nu}_\ell$ underlies a plethora of (semi)leptonic hadron decays. Ref.~\cite{Gonzalez-Alonso:2016etj} studied $d(s)\to u \ell \bar{\nu}_\ell$ transitions, such as nuclear, baryon and meson decays, within the SMEFT framework and obtained bounds for 14 combinations of effective low-energy couplings between light quarks and leptons ($\epsilon_i^{d_I e_J}$). Ignoring the CKM mixing at $\cO(\Lambda^{-2})$, the effective couplings of strange quarks depend only on flavor-off-diagonal Wilson coefficients (see \aref{LEFFE}). Marginalizing over them, we obtain the likelihood for 6 combinations of effective couplings together with the $\tilde V_{ud}$ CKM parameter:\footnote{
There is a small (but nonzero) correlation with the effective couplings of strange quarks that we marginalized over. This must be taken into account when going to specific scenarios. The full likelihood is available in Ref.~\cite{Gonzalez-Alonso:2016etj}.
}
\bea
\label{eq:6Dleffe}
\left(
\begin{array}{c}
\tilde{V}_{ud} \\
\Delta_{\rm CKM} \\
 \epsilon_R^{de}\\
 \epsilon_S^{de} \\
  \epsilon_P^{de} \\
\eT^{de} \\
\Delta^d_{LP}
\end{array}
\right)
=
\left(
\begin{array}{c}
 0.97451(38) \\
 -(1.2\pm 8.4)\cdot 10^{-4} \\
  -(1.3\pm 1.7)\cdot 10^{-2}  \\
    (1.4\pm 1.3) \cdot 10^{-3} \\
 (4.0\pm 7.8) \cdot 10^{-6} \\
  (1.0\pm 8.0) \cdot 10^{-4} \\
 (1.9\pm 3.8)\cdot 10^{-2} 
\end{array}
\right)\,,~
\rho =
\left(
\scriptsize{
\begin{array}{ccccccc}
 1. & 0.88 & 0. & 0.82 & 0.01 & 0. & 0.01 \\
 0.88 & 1. & 0. & 0.73 & 0.01 & 0. & 0.01 \\
 0. & 0. & 1. & 0. & -0.87 & 0. & -0.87 \\
 0.82 & 0.73 & 0. & 1. & 0.01 & 0. & 0.01 \\
 0.01 & 0.01 & -0.87 & 0.01 & 1. & 0. & 0.9995 \\
 0. & 0. & 0. & 0. & 0. & 1. & 0. \\
 0.01 & 0.01 & -0.87 & 0.01 & 1. & 0. & 1. \\
\end{array}
}\right)\,,
\eea
in the $\overline{MS}$ scheme at $\mu=2$ GeV. The effective couplings $\epsilon$ were defined in \sref{CC}, and $\Delta^d_{LP} \approx \eL^{de}- \eL^{d\mu} + 24  \epsilon_P^{d\mu}$. 
See \aref{LEFFE} for the complete likelihood~\cite{Gonzalez-Alonso:2016etj} that also involves the effective couplings  of the strange quarks and allows one to constrain some off-diagonal Wilson coefficients. 
Using \eref{RGEepsilon} we can run these results up to the weak scale, where the matching with the SMEFT is carried out, \emph{cf.} \eref{epsilon} and \eref{deltackm}.

It is useful to recall the physics behind these bounds~\cite{Gonzalez-Alonso:2016etj}. Roughly speaking, $\tilde{V}_{ud}$ and $\epsilon^{de}_{R,S,P,T}$ were obtained comparing the total rates of various superallowed nuclear decays and $\pi\to e\nu_e$, as well as using various differential distributions in $\pi\to e\nu\gamma$ and neutron decay. The comparison with $\Gamma(\pi\to \mu\nu_\mu)$ provides us with $\Delta^d_{LP}$, and the combination of the obtained $\tilde{V}_{ud}$ with $V_{us}$, extracted from (semi)leptonic kaon decays, makes possible to extract $\Delta_{CKM}$.  

%/////////////////////////////////////////////////////////////////////////////////////////////////////////
\subsection{Quark pair production in $e^+ e^-$ collisions}
\label{sec:epem}

Electron-positron colliders operating at center-of-mass energies above or below the $Z$ mass provide complementary information about 4-fermion operators containing electrons. 
Unlike the low-energy experiments discussed above, they also probe flavor-conserving operators with strange, charm and bottom quarks. Typically, the experiments quote the total measured cross section for 
$\sigma_{q} \equiv \sigma(e^+ e^- \to  q \bar q)$ and the asymmetry $A_{FB}^{q} =  {\sigma_q^{\rm FB} \over \sigma_{q}}$,  where  $\sigma_{q}^{FB}$ is the difference between the cross sections with the electron going forward and backward in the center-of-mass frame. 
In the presence of dimension-6 operators, at $\cO(\Lambda^{-2})$ these cross sections are modified as follows 
\bea 
\label{eq:EPEM_sigma}
\delta\sigma_q   & = & \frac{1}{8\pi s} 
\left[- e^2 (g_L^2+g_Y^2) \frac{s}{s-m_Z^2}\left(\delta A_{Fq}+\delta A_{Bq}\right)
+(g_L^2+g_Y^2)^2 \frac{s^2}{(s-m_Z^2)^2}\left(\delta B_{Fq}+\delta B_{Bq}\right)\right]
\nnl 
&+&  \frac{1}{8\pi v^2} (g_L^2+g_Y^2) \frac{s}{s-m_Z^2}
\left(\hat g_L^{Ze} \hat g_L^{Zq} c_{LL}+ \hat g_L^{Ze} \hat g_R^{Zq} c_{LR}+ \hat g_R^{Ze} \hat g_L^{Zq} c_{RL}+ \hat g_R^{Ze} \hat g_R^{Zq} c_{RR}\right)\
\nnl 
&- & \frac{1}{8\pi v^2} e^2 Q_q \left(c_{LL}+c_{LR}+c_{RL}+c_{RR}\right), 
\eea 
\bea 
\label{eq:EPEM_sigmafb}
\delta \sigma_q^{\rm FB} &=& \frac{3}{32\pi s} \left[- e^2 (g_L^2+g_Y^2) \frac{s}{s-m_Z^2}\left(\delta A_{Fq}-\delta A_{Bq}\right)+(g_L^2+g_Y^2)^2 \frac{s^2}{(s-m_Z^2)^2}\left(\delta B_{Fq}-\delta B_{Bq}\right)\right]
\nnl  &+& 
\frac{3}{32\pi v^2}(g_L^2+g_Y^2) \frac{s}{s-m_Z^2}
\left(\hat  g_L^{Ze} \hat g_L^{Zq} c_{LL}+ \hat  g_R^{Ze}\hat  g_R^{Zq} c_{RR}- \hat  g_L^{Ze} \hat  g_R^{Zq} c_{LR}- \hat  g_R^{Ze} \hat  g_L^{Zq} c_{RL}\right) 
\nnl &-& 
\frac{3}{32\pi v^2}e^2 Q_q \left(c_{LL}+c_{RR}-c_{LR}-c_{RL} \right ), 
\eea
where $\sqrt{s}$ is the center-of-mass energy of the $e^+e^-$ collision, 
$ \hat g^{Zf} \equiv T_f^3 - s_\theta^2 Q_f$ (i.e., the SM values),  and 
\begin{eqnarray}
\delta A_{Fq}&=& Q_q \left(
 \delta  g_L^{Ze} \hat g_L^{Zq} +  \delta g_R^{Ze} \hat g_R^{Zq} 
 + \hat g_L^{Ze} \delta g_L^{Zq} + \hat g_R^{Ze} \delta g_R^{Zq} \right),  \\ 
\delta A_{Bq}&=& Q_q \left( \delta  g_L^{Ze} \hat g_R^{Zq}+  \delta g_R^{Ze}  \hat  g_L^{Zq} 
+  \hat  g_L^{Ze}  \delta g_R^{Zq}+  \hat  g_R^{Ze} \delta g_L^{Zq}   \right), 
\nnl 
\delta B_{Fq}&=& \hat  g_L^{Ze}\left(\hat  g_L^{Zq}\right)^2 \delta  g_L^{Ze} 
+ \hat  g_R^{Ze} \left(\hat  g_R^{Zq}\right)^2 \delta g_R^{Ze} 
 + \left(\hat  g_L^{Ze}\right)^2 \hat  g_L^{Zq}  \delta g_L^{Zq} 
  +  \left(\hat  g_R^{Ze}\right)^2 \hat  g_R^{Zq}  \delta g_R^{Zq}, 
  \nonumber \\
\delta B_{Bq}&=& \hat  g_L^{Ze}\left(\hat  g_R^{Zq}\right)^2\delta  g_L^{Ze}
 +  \hat  g_R^{Ze} \left(\hat  g_L^{Zq}\right)^2 \delta g_R^{Ze}
 + \left(\hat  g_R^{Ze}\right)^2 \hat  g_L^{Zq}  \delta g_L^{Zq}  +  \left(\hat  g_L^{Ze}\right)^2 \hat  g_R^{Zq}  \delta g_R^{Zq}. \nonumber
\end{eqnarray}
For the up-type quark production, $q = u_J$,  the four-fermion Wilson coefficients $c_{XY}$ in \eref{EPEM_sigma} and \eref{EPEM_sigmafb} are given by 
\beq 
c_{LL} =  [c_{\ell q}]_{11JJ } -[c^{(3)}_{\ell q}]_{11JJ}, \quad 
c_{LR} = [c_{\ell u}]_{11JJ}, \quad 
c_{RL} =  [c_{eq}]_{11JJ}, \quad 
c_{RR} =  [c_{eu}]_{11JJ}, 
\eeq 
while for  the down-type quark production, $q = d_J$, 
\beq 
c_{LL} =  [c_{\ell q}]_{11JJ } + [c^{(3)}_{\ell q}]_{11JJ}, \quad 
c_{LR} = [c_{\ell d}]_{11JJ}, \quad 
c_{RL} =  [c_{eq}]_{11JJ}, \quad 
c_{RR} =  [c_{ed}]_{11JJ}. 
\eeq 
The operators $O_{\ell equ}$, $O^{(3)}_{\ell equ}$ and  $O_{\ell eqd}$ do not enter at $\cO(\Lambda^{-2})$ because they do not interfere with the SM amplitudes due to the different chirality structure.

The LEP-2 experiment studied  $e^+ e^-$ collisions at energies above the $Z$-pole, ranging from $\sqrt s =$ 130 Gev to $\sqrt s =$ 209 GeV. 
Available data includes the total cross section $\sigma(q\bar{q})\equiv \sum_{q=u,d,s,c,b} \sigma_q$ \cite{Schael:2013ita}, 
as well as the total cross section and forward-backward asymmetry for the charm and for the bottom quark production \cite{Alcaraz:2006mx}. 
This amounts to 5 distinct observables, each measured at different $\sqrt{s}$. 
From \eref{EPEM_sigma} and \eref{EPEM_sigmafb}, given the energy dependence, each of these observables should resolve 4 different combinations of dimension-6 Wilson coefficients.\footnote{Note that two of these combinations involve only vertex corrections though.} 
In practice, the energy range scanned by LEP-2 is not large enough to efficiently disentangle these different combinations. 
Therefore, in our fit we also include earlier, less precise measurements of heavy quark production below the Z-pole. 
Specifically, we include   the measurements  from the VENUS \cite{Abe:1993xr} and TOPAZ \cite{Inoue:2000hc} collaborations of  the $c \bar c$ and $b \bar b$ pair production at  $\sqrt{s}=58$~GeV (total cross sections and FB asymmetries).

%//////////////////////////////////////////////////////////////////////////////////////////////////
\subsection{Other measurements}

To increase the power of our global analysis, in this section we will combine the observables described above  with those considered previously in Refs.~\cite{Efrati:2015eaa,Falkowski:2015krw}.
At this point there are more parameters than observables,  hence more experimental input is needed.  
The SMEFT corrections to low-energy observables typically depend on linear combinations of 4-fermion Wilson coefficients and vertex corrections $\delta g$. 
The latter can be independently constrained by the so-called pole observables where a single W or Z boson is  on-shell. 
We use the set of pole observables described in Ref.~\cite{Efrati:2015eaa}. 
As advertised in that reference, all diagonal $\delta g$ can be simultaneously constrained with a very good precision.\footnote{The observables in Ref.~\cite{Efrati:2015eaa} do not constrain $\delta g^{Zt}_R$, which  is however not needed in our analysis.} 
Moreover, we use the low-energy and $e^+ e^-$ collider observables probing 4-lepton operators. 
Our analysis closely resembles that in Ref.~\cite{Falkowski:2015krw} with the following  differences:
\ben
\item Instead of combining ourselves the results of different experiments measuring the scattering of muon neutrinos on electrons, we use the PDG combination for the low-energy $\nu_\mu$-$e$ couplings from Table~10.8 of Ref.~\cite{Olive:2016xmw}:
\beq
g_{LV}^{\nu_\mu e} =  -0.040 \pm 0.015, 
\qquad g_{LA}^{\nu_\mu e} = -0.507 \pm 0.014, 
\eeq  
with the correlation coefficient $\rho = -0.05$.
\item  Instead of recasting the weak mixing angle measured in parity-violating electron scattering \cite{Anthony:2005pm}, we use the  PDG result for the parity-violating  effective self-coupling of electrons \cite{Olive:2016xmw}:
\beq
g_{AV}^{ee} = 0.0190 \pm 0.0027. 
\eeq 
\item To evaluate SMEFT corrections to $e^+ e^-$ collider observables we use the electroweak couplings at the scale $m_Z$ (instead of $200$~GeV). 
\item We add the measurement of the $\tau$ polarization ${\cal P}_{\tau}$ and its FB asymmetry $A_{\cal P}$ in $e^+ e^- \to \tau^+ \tau^-$ production at $\sqrt{s}=58$~GeV by the VENUS collaboration~\cite{Hanai:1997ty}:  
\beq
{\cal P}_{\tau} = 0.012 \pm 0.058, \qquad A_{\cal P}= 0.029 \pm 0.057.  
\eeq 
The analytic expressions for ${\cal P}_{\tau}$ and $A_{\cal P}$ in function of the SMEFT parameters and $\sqrt{s}$ are easy to obtain but are too long to be quoted here. 
Instead, we give the numerical expressions at   $\sqrt{s}=58$~GeV: 
\bea
\delta {\cal P}_{\tau} & \approx &   
 - 0.87 \delta g^{Ze}_L -  0.93\delta g^{Ze}_R + 0.17\delta g^{Z\tau}_L + 0.25 \delta g^{Z\tau}_R
\nnl  
& + & 0.21 [c_{ee}]_{1133} + 0.32 [c_{le}]_{1133} - 0.34 [c_{le}]_{3311} 
-  0.20 ([c_{\ell \ell}]_{1133} + [c_{\ell \ell}]_{1331}), 
\nnl 
\delta A_{\cal P} & \approx &   
 0.13 \delta g^{Ze}_L + 0.19\delta g^{Ze}_R -0.65\delta g^{Z\tau}_L - 0.70 \delta g^{Z\tau}_R
\nnl  
& + & 0.16 [c_{ee}]_{1133} - 0.25 [c_{le}]_{1133} + 0.24 [c_{le}]_{3311} 
-  0.15 ([c_{\ell \ell}]_{1133} + [c_{\ell \ell}]_{1331}).
\eea
\item We  include the constraints from the trident production $\nu_\mu \gamma^* \to \nu_\mu \mu^+ \mu^-$~\cite{Geiregat:1990gz,Mishra:1991bv,Altmannshofer:2014pba}.
Dimension-6 operators modify the trident cross section as
\bea
{\sigma_{\rm trident} \over \sigma_{\rm trident, \, SM}} 
& \approx & 
1 
+ 2{g_{LV}^{\nu_\mu \mu,\rm SM} \delta g_{LV}^{\nu_\mu \mu} + g_{LA}^{\nu_\mu \mu,\rm SM} \delta g_{LA}^{\nu_\mu \mu}
\over 
(g_{LV}^{\nu_\mu \mu,\rm SM})^2 + (g_{LA}^{\nu_\mu \mu,\rm SM})^2}.
\eea 
\een 
The first 3 modifications lead to negligible numerical differences compared to the fit in Ref.~\cite{Falkowski:2015krw}. 
The 4th one allows us to break the degeneracy between $[c_{\ell e}]_{1133}$ and $[c_{\ell e}]_{3311}$ and improve constraints on other 4-lepton operators involving $\tau$. 
The last modification leads to a constraint on  one linear combination of 4-muon dimension-6 operators.

%%%%%%%%%%%%%%%%%%%%%%%%%%%%%%%

 \begin{table}
 \begin{center}
 \begin{tabular}{|c|c|c|c|c|}
\hline
{\color{blue}{Class}} 			& {\color{blue}{Observable}}	& {\color{blue}{Exp. value}}   &   {\color{blue}{Ref. \& Comments}}   &  {\color{blue}{SM value}} 
  \\  \hline   \hline
$\nu_e\nu_eqq$					&	$R_{\nu_e \bar \nu_e}$	&	$0.41(14)$		&	CHARM~\cite{Dorenbosch:1986tb}			&	$0.33$
  \\ \hline \hline
\multirow{4}{*}{$\nu_\mu\nu_\mu qq$}
								&	$(g_L^{\nu_\mu})^2$ 	&	$0.3005(28)$			&	\multirow{4}{*}{PDG~\cite{Olive:2016xmw}, $\rho\approx1$}	&	$0.3034$
   \\  \cline{2-3}\cline{5-5}
								&	$(g_R^{\nu_\mu})^2$ 	&	$0.0329(30)$			& 									&     $0.0302$
   \\  \cline{2-3}\cline{5-5}
								&	$\theta_L^{\nu_\mu}$ 	&	$2.500(35)$				&									&     $ 2.4631$
   \\  \cline{2-3}\cline{5-5}
								&	$\theta_R^{\nu_\mu}$ 	&	$4.56^{+0.42}_{-0.27}$	& 									&     $5.1765$
 \\ \hline \hline
\multirow{5}{*}{\begin{minipage}{0.7in}PV low-E $~~~~ee qq$\end{minipage}}
								&	$g^{eu}_{AV} + 2g^{ed}_{AV}$ 	&	$0.489(5)$		& 	\multirow{3}{*}{PDG~\cite{Olive:2016xmw}, $\rho\neq 1$}					&     $0.4951$
   \\  \cline{2-3}\cline{5-5}
								&	$2 g^{eu}_{AV} -  g^{ed}_{AV}$ 	&	$-0.708(16)$	& 									&     $-0.7192$
   \\  \cline{2-3}\cline{5-5}
								&	$2 g^{eu}_{VA} -  g^{ed}_{VA}$ 	&	$-0.144(68)$	&									&     $-0.0949$
   \\  \cline{2-5}
								&	\multirow{2}{*}{$g^{eu}_{VA}-g^{ed}_{VA}$} 	&$-0.042(57)$		&\multirow{2}{*}{SAMPLE~\cite{Beise:2004py}}	&	\multirow{2}{*}{$-0.0627$}
	\\
								&										&	$-0.120(74)$	&			&	
  \\ \hline \hline
\multirow{2}{*}{\begin{minipage}{0.7in}PV low-E $~~~~\mu\mu qq$\end{minipage}}
								&	$b_{\rm SPS} (\lambda = 0.81)$ 	&	$-1.47(42) \cdot 10^{-4}$	& 	\multirow{2}{*}{SPS~\cite{Argento:1982tq}}			&     $-1.56 \cdot 10^{-4}$
   \\  \cline{2-3}\cline{5-5}
								&	$b_{\rm SPS} (\lambda = 0.66)$ 	&	$-1.74(81) \cdot 10^{-4}$	&						&     $-1.57 \cdot 10^{-4}$
   \\ \hline \hline
$d(s)\to u \ell\nu$				&	$\epsilon_i^{d_j \ell}$			&	\eref{6Dleffe}			&	Ref.~\cite{Gonzalez-Alonso:2016etj} 	&	0
   \\ \hline \hline
\multirow{3}{*}{$e^+e^-\to q\bar{q}$}
	&	$\sigma(q\bar{q})$		&					&	LEPEWWG~\cite{Schael:2013ita}, $\rho\neq 1$	&
\\  \cline{2-2}\cline{4-4}
	&	$\sigma_c, \sigma_b$	&	$f(\sqrt{s})$	&	\multirow{2}{*}{\begin{minipage}{1.9in}~~~~~~LEPEWWG~\cite{LEP:2003aa},\\VENUS~\cite{Abe:1993xr}, TOPAZ~\cite{Inoue:2000hc}\end{minipage}}	&	$f(\sqrt{s})$
\\  \cline{2-2}
	&	$A_{FB}^{cc},A_{FB}^{bb}$	&				&												&
  \\ \hline \hline
\multirow{2}{*}{$\nu_\mu\nu_\mu e e$}
								&	$g_{LV}^{\nu_\mu  e}$ 	&	$ -0.040(15)$			&	\multirow{2}{*}{PDG~\cite{Olive:2016xmw}, $\rho\neq 1$}	&	$ -0.0396$
   \\  \cline{2-3}\cline{5-5} 
								&	$g_{LA}^{\nu_\mu  e}$ 	&	$ -0.507(14)$			& 									&     $-0.5064$
\\ \hline \hline
$e^- e^-  \to e^- e^-$ & $g_{AV}^{ee}$  &  $0.0190(27)$ &  PDG~\cite{Olive:2016xmw}  & $0.0225$
\\ \hline \hline
\multirow{2}{*}{$\nu_\mu\gamma^*\!\to\! \nu_\mu \mu^+ \mu^-$}	& \multirow{2}{*}{${\sigma\over\sigma_{\rm SM}}$}	&$1.58(57)$	&CHARM~\cite{Geiregat:1990gz}	&\multirow{2}{*}{$1$}
\\
														&										&$0.82(28)$	&CCFR~\cite{Mishra:1991bv}		&			  \\ \hline \hline  
\multirow{2}{*}{$\tau \to \ell \nu \nu$}
								&	$G_{\tau e}^2/G_F^2$ 	&	$1.0029(46)$			&	\multirow{2}{*}{PDG~\cite{Olive:2016xmw}, 
								% PSI~\cite{Danneberg:2005xv},  
								$\rho\approx1$}	&	$1$
   \\  \cline{2-3}\cline{5-5}
								&	$G_{\tau \mu}^2/G_F^2$ 	&	$0.981(18)$			& 									&     $1$
%   \\  \cline{2-3}\cline{5-5} &	Michel~$\eta$ 	&	$-0.0021(71)$				&								 &     $0$
   %\\  \cline{2-3}\cline{5-5} &	Michel~$\beta'/A$ 	&	$-0.0013(36)$	& 									&     $0$
%
\\ \hline \hline
\multirow{3}{*}{$e^+e^-\to \ell^+ \ell^-$}
	&	${d \sigma(ee) \over d \cos \theta }$		&					&	LEPEWWG~\cite{Schael:2013ita}, $\rho \approx 1$	&
\\  \cline{2-2}\cline{4-4}
	&	$\sigma_\mu, \sigma_\tau, {\cal P}_\tau$	&	$f(\sqrt{s})$	&	\multirow{2}{*}{\begin{minipage}{1.9in}~~~~~~LEPEWWG~\cite{LEP:2003aa},\\  \phantom {aaaa} VENUS~\cite{Hanai:1997ty} \end{minipage}}	&	$f(\sqrt{s})$
\\  \cline{2-2}
	&	$A_{FB}^{\mu},A_{FB}^{\tau}$	&				&												&
  \\ \hline %\hline
 \end{tabular}
\end{center}
\caption{
Summary of experimental input (sensitive to LLQQ and LLLL contact interactions) used in our fit. The correlations that are taken into account in our fit are specified. Each observable in $e^+e^-\to f\bar{f}$ is measured at various c.o.m. energies, which we denote in the table by $f(\sqrt{s})$. The specific numerical values can be found in the corresponding original references. We also use the set of pole observables described in Ref.~\cite{Efrati:2015eaa} in order to independently  constrain the vertex corrections $\delta g$. }
\label{tab:inputsummary}
 \end{table}

%%%%%%%%%%%%%%%%%%%%%%%%%%%%%%%
\section{Global Fit}
\label{sec:gf}
\setcounter{equation}{0} 

%-----------------------------------------------
\subsection{Scope}
\label{sec:GF_scope}

The main goal of this paper is to provide model-independent constraints on flavor-diagonal 2-lepton-2-quark operators summarized in \tref{2l2q}.  
 Among  the chirality-conserving ones, most of the observables considered in this paper probe the operators involving the 1st generation leptons. 
There are 21 such operators and for an easy reference we list here their Wilson coefficients: 
\beq 
\label{eq:list_11JJ}
~~~ [c_{\ell q}]_{11JJ}, \  [c_{\ell q}^{(3)}]_{11JJ}, \  [c_{\ell u}]_{11JJ}, \   [c_{\ell d}]_{11JJ}, \   [c_{eq}]_{11JJ}, \  [c_{eu}]_{11JJ}, \  [c_{ed}]_{11JJ} \ , 
 \qquad J=1,2,3.
 \eeq 
Scattering of muons and muon neutrinos on nucleons gives us access to chirality-conserving operators involving 2nd generation leptons and 1st generation quarks. 
There are 7 such operators: 
\beq 
\label{eq:list_2211}
[c_{\ell q}]_{2211}, \ [c^{(3)}_{\ell q}]_{2211}, \  [c_{\ell u}]_{2211}, \   [c_{\ell d}]_{2211}, \   [c_{eq}]_{2211}, \  [c_{eu}]_{2211}, \  [c_{ed}]_{2211}. 
\eeq 
Finally, the likelihood in \eref{6Dleffe} summarizing the constraints from low-energy flavor observables gives us also access to chirality-violating operators involving 1st and 2nd generation leptons and and 1st generation quarks. 
There are 6 such operators: 
 \beq 
\label{eq:list_hv}
 [c_{lequ}]_{JJ11}\ , \  [c_{ledq}]_{JJ11}\ ,  \ [c^{(3)}_{lequ}]_{JJ11} \ , 
 \qquad J=1,2~,
\eeq 
which should be understood as evaluated at the renormalization scale $\mu=m_Z$ unless otherwise stated.

We will use the observables summarized in \sref{lee} to constrain as many as possible of the 34 Wilson coefficients in Eqs.~(\ref{eq:list_11JJ})-(\ref{eq:list_hv}). 
We will also present simultaneous constraints on these parameter, together with the vertex corrections  and 4-lepton Wilson coefficients.

%-----------------------------------------------
\subsection{Flat directions}
\label{sec:GF_flat}

Not all linear combinations of the parameters Eqs.~(\ref{eq:list_11JJ})-(\ref{eq:list_hv}) can be constrained by the observables we consider. 
Before venturing into a global fit, we need to count the independent constraints and determine the flat directions in the parameter space.
In \tref{inputsummary} we have the following probes of  LLQQ operators: 
\bi
\item 1 combination of the parameters in \eref{list_11JJ} is constrained (poorly)  via the only $\nu_e\nu_e qq$ measurement ($R_{\nu_e \bar \nu_e}$);
\item 4 combinations in \eref{list_2211} are constrained via $\nu_\mu\nu_\mu qq$ measurements;
\item 4 new combinations in \eref{list_11JJ}  are constrained via PV low-energy $eeqq$ measurements ($g_{VA/AV}^{eq}$);
\item 1 different combination in \eref{list_2211} is constrained (poorly) via PV low-energy $\mu\mu qq$ measurements ($b_{\rm SPS}$), which also probe a second combination already constrained by $\nu_\mu\nu_\mu qq$ data;
\item  5 additional combinations in Eqs.~(\ref{eq:list_11JJ})-(\ref{eq:list_hv}) are constrained by low-energy flavor observables ($d(s)\to u \ell\nu_\ell$ transitions);\footnote{%
The likelihood in \eref{6Dleffe} also independently constrains $\delta g^{W q_1}_R$.}
\item 10 additional combinations in \eref{list_11JJ} are probed by $e^+ e^-\to q\bar{q}$ data, through the measurement of the the total hadronic cross section and heavy flavor ($b$ and $c$) fractions and asymmetries.
\ei 
All together we have 25  constraints on 34 parameters, which leaves 9 flat directions. 
These  can be characterized quite concisely:
\begin{align}
\label{eq:GF_eflat}
& ({\rm \bf F1}): \quad   [c_{\ell u}]_{1133},  \quad ({\rm \bf  F2}): \quad  [c_{eu}]_{1133}, \quad ({\rm  \bf  F3}): \quad  [c_{\ell q}^{(3)}]_{1133} = -  [c_{\ell q}]_{1133}, 
\nnl 
& ({\rm  \bf  F4}): \quad   [c_{\ell q}^{(3)}]_{1122} =   [c_{\ell q}]_{1122}, \quad   [c_{\ell d}]_{1122} = \left (-5 + {3 g_L^2 \over g_Y^2} \right ) [c_{\ell q}]_{1122}, \quad 
  [c_{e d}]_{1122} = \left (3 -  {3 g_L^2 \over g_Y^2} \right ) [c_{\ell q}]_{1122}, 
  \nnl 
& ({\rm  \bf  F5}): \quad  [c_{\ell q}]_{1111} = - [c_{\ell u}]_{1111} = - [c_{\ell d}]_{1111} = - [c_{e q}]_{1111} 
= [c_{e u}]_{1111}  =   [c_{e d}]_{1111}, 
\nnl 
&  ({\rm  \bf  F6}): \quad  [c_{eq}]_{2211}  = - [c_{ed}]_{2211}, 
 \quad  ({\rm  \bf  F7 }): \quad  [c_{eq}]_{2211}  = 2 [c_{eu}]_{2211}, 
 \nnl 
&  ({\rm  \bf  F8,F9}): \quad 
 0.86 [c_{ledq}]_{2211} - 0.86 [c_{lequ}]_{2211} + 0.012  [c_{ledq}^{(3)}]_{2211} = 0. 
\end{align}
The flat directions F1, F2, F3 arise because low-energy precision measurements do not probe the top quark couplings, which may be amended one day by  $e^+ e^-$ collider operating above the $t \bar t$ threshold.
F4 is due to the insufficient information about the strange quark couplings, and it could be lifted by off Z-pole measurements of the strange asymmetry. 
F5 is the consequence of the fact that  the parity conserving operator $(\bar e \gamma_\mu \gamma_5 e) \sum_q (\bar q \gamma_\mu \gamma_5 q)$ and the axial neutrino-quark interaction  $(\bar \nu_L \gamma_\mu \nu_L)\sum_q (\bar q \gamma_\mu \gamma_5 q)$ are unconstrained by low-energy measurements and by $e^+e^-$ colliders.
F6 and F7 are due to little data on muon scattering on nucleons. 
Finally, F8 and F9 appear because, with our approximations, the low-energy flavor observables probe only one combination  of light quark couplings to muons (through $\pi\to\mu\nu$).  The low-energy constraint on $\eps_P^{d \mu} =  ([c_{ledq}]_{2211} -  [c_{lequ}]_{2211} )/2$ at $\mu=2$~GeV (via $\Delta^d_{LP}$ in \eref{6Dleffe}), after taking into account the running up to $m_Z$,  becomes a constraint on the linear combination  in the last line of \eref{GF_eflat}.

In order to isolate the flat directions we define
\bea 
  \, [\hat c_{e q}]_{1111}  & = &   [c_{e q}]_{1111}   +  [c_{\ell q}]_{1111}, 
     \nnl 
  \, [\hat c_{\ell u}]_{1111}  & = &      [c_{\ell u}]_{1111} +  [c_{\ell q}]_{1111}  -   [\hat c_{e q}]_{1111}, 
     \nnl 
   \,   [\hat c_{\ell d}]_{1111}   & = &   [c_{\ell d}]_{1111}+  [c_{\ell q}]_{1111}   -   [\hat c_{e q}]_{1111} , 
   \nnl
     \,   [\hat c_{eu}]_{1111}  & = &     [c_{eu}]_{1111} -   [c_{\ell q}]_{1111} ,
       \nnl 
    \,  [\hat c_{ed}]_{1111}   & = &  [c_{ed}]_{1111} -   [c_{\ell q}]_{1111} ,
     \nnl 
  \,  [\hat c_{\ell q}^{(3)}]_{1122}   & = &      [c_{\ell q}^{(3)}]_{1122} -  [c_{\ell q}]_{1122},
     \nnl 
   \,  [\hat c_{\ell d}]_{1122}  & = &    [c_{\ell d}]_{1122}  
   +   \left (5 -  {3 g_L^2 \over g_Y^2} \right ) [c_{\ell q}]_{1122}  - [\hat c_{e q}]_{1111}  ,
     \nnl 
   \,  [\hat c_{ed}]_{1122}   & = &      [c_{ed}]_{1122}  
     -  \left (3 -  {3 g_L^2 \over g_Y^2} \right )[c_{\ell q}]_{1122}  - [\hat c_{e q}]_{1111}  ,
\nonumber
 \eea 
\bea 
\label{eq:chats}
 \,  [\hat c_{\ell q}^{(3)}]_{1133}   & = &    [c_{\ell q}^{(3)}]_{1133}  +  [c_{\ell q}]_{1133}, 
 \nnl
 \, [\hat c_{eq}]_{2211}  & =&    [c_{eq}]_{2211} + [c_{ed}]_{2211} - 2 [c_{eu}]_{2211},
 \nnl 
 \eps_P^{d \mu} (2~\rm{GeV})  &= & 0.86 [c_{ledq}]_{2211} - 0.86 [c_{lequ}]_{2211} + 0.012  [c_{ledq}^{(3)}]_{2211}, 
 \nnl 
\, [\hat c_{\ell \ell}]_{2222} &= &  [c_{\ell \ell}]_{2222} 
 + {2 g_Y^2 \over g_L^2 + 3 g_Y^2} [c_{\ell e}]_{2222}.  
\eea 
The last variable projects out the flat direction among 4-muon operators in the trident observable.
Using these variables, the global likelihood depends on the Wilson coefficients 
on the right-hand sides of Eqs.~(\ref{eq:chats}) only via the $\hat c$ and $\eps_P^{d \mu}(2~\rm{GeV})$ combinations.\footnote{Let us stress that the LLQQ coefficients in the r.h.s. of $\eps_P^{d \mu}(2~\rm{GeV})$ in \eref{chats} are defined at $\mu=m_Z$.}  
Moreover, the dependence on  $[\hat c_{e q}]_{1111}$ appears only thanks to the loose $R_{\nu_e \bar \nu_e}$ constraint,  and thus we know beforehand that there is  no sensitivity  to  $[\hat c_{e q}]_{1111} \lesssim 1$.   

%---------------------------------------------------
\subsection{Reconnaissance} 
\label{sec:1-by-1-fit}
We start by presenting the constraints in the case when only one of the LLQQ operators is present at a time, and all vertex corrections and 4-lepton operators vanish. 
We stress that this is just a warm-up exercise and not our main result. 
Indeed, one-by-one constraints are basis dependent and could be different if another basis of dimension-6 operators was used. 
Only the global likelihood encoding the correlated constraints on all Wilson coefficients in a given basis has a model-independent meaning. 
The main  purpose of this exercise is to compare the sensitivity of various experiments to a few particular directions in the space of Wilson coefficients. 

\begin{table}
\bc
%\scriptsize 
$(ee)(qq)$
\footnotesize
\begin{tabular}{|c|c|c|c|c|c|c|c|c}
\hline  
&  $ [c_{\ell q}^{(3)}]_{1111} $ &  $[c_{\ell q}]_{1111}$   & $[c_{\ell u}]_{1111}$  & $[c_{\ell d }]_{1111}$ 
& $[c_{eq}]_{1111}$  & $[c_{eu}]_{1111}$ &   $[c_{ed}]_{1111}$  \\ 
\hline 
CHARM
& $-80  \pm 180 $ & $700 \pm  1800$ & $370 \pm 880$ &  $-700 \pm 1800$ & x & x & x 
\\ \hline 
APV
&  $27 \pm 19 $ &  $\bfblue{ 1.6 \pm 1.1} $ &  $\bfblue{ 3.4 \pm 2.3}$ &  $\bfblue{ 3.0 \pm 2.0}$ &  $\bfblue{ -1.6 \pm 1.1 }$ &  $\bfblue{ -3.4 \pm 2.3 }$ &  $\bfblue{ -3.0 \pm 2.0}$ 
\\ \hline  
QWEAK 
& $7.0 \pm 12 $ & $-2.3 \pm 4.0 $ & $-3.5 \pm 6.0 $ & $-7  \pm 12$ & $2.3 \pm 4.0$ & $3.5 \pm  6.0 $ & $7 \pm 12$ 
\\ \hline
PVDIS 
& $-8 \pm 12  $ & $24  \pm  35 $ & $38  \pm 48 $ & $-77 \pm  96 $ & $-77 \pm 96 $ & $-12 \pm 17 $ & $24 \pm  35$ 
\\ \hline
SAMPLE 
& $-8 \pm 45 $ & x  & $ -17 \pm 90$  &   $17 \pm  90 $ & x & $-17 \pm 90$  &   $17 \pm  90 $ 
 \\ \hline 
$d_j\to u \ell\nu$ & $\bfblue{ 0.38 \pm 0.28}$ & x & x & x & x & x & x 
 \\ \hline 
LEP-2
&  $3.5 \pm 2.2 $ &$ - 42 \pm 28 $ & $-21 \pm 14$  & $42 \pm 28$  
& $-18 \pm 11$ & $-9.0 \pm 5.7$ & $18 \pm 11$		\\			
\hline 
\end{tabular}
\ec
\bc 
$(\mu \mu)(qq)$
\begin{tabular}{|c|c|c|c|c|c|c|c|}
\hline
&  $ [c_{\ell q}^{(3)}]_{2211} $ &  $[c_{\ell q}]_{2211}$   & $[c_{\ell u}]_{2211}$  & $[c_{\ell d }]_{2211}$  
& $[c_{eq}]_{2211}$  & $[c_{eu}]_{2211}$ &   $[c_{ed}]_{2211}$\\   
\hline 
PDG $\nu_\mu$ & $ 20 \pm 15$ & ${\bfblue{4 \pm 21}}$ & $\bfblue{18 \pm 19}$ & $\bfblue{-20 \pm 37}$ & x & x & x   \\
\hline
SPS & $0 \pm 1000$ & $0 \pm 3000$ & $0 \pm 1500$ & $0  \pm 3000$ & $\bfblue{40 \pm 390}$ & $\bfblue{-20 \pm 190}$ & $\bfblue {40 \pm 390}$    \\
\hline
$d_j\to u \ell\nu$ & $\bfblue{ -0.4 \pm 1.2} $ & x & x & x  & x & x &x  \\
\hline 
\end{tabular}
\ec 
\caption{68\% C.L. constraints (in units of $10^{-3}$) on chirality-conserving $(ee)(qq)$ and $(\mu \mu)(qq)$  operators from different precision experiments.
The bounds are derived assuming that only one operator is present at a time.
See \tref{inputsummary} and main text for further details about the different experiments. 
The best constraint in each case is highlighted in blue, while `x' signals that the operator is not probed at tree level by that experiment or combination.}
\label{tab:1by1}
\end{table}

The one-by-one constraints on chirality-conserving LLQQ operators involving 1st generation quarks are shown in \tref{1by1}.
One can see that atomic parity violation is the most sensitive probe for most of the operators with electrons and the first generation quarks.
The exception is $[O_{\ell q}^{(3)}]_{1111}$, which contributes to charged-current transitions and can be probed in $d\to ue\bar{\nu}_e$ decays.\footnote{The single-operator bounds from $d(s)\to u\ell\bar{\nu}_\ell$ data shown in this section are obtained using the likelihood of \eref{6Dleffe}, which was marginalized over strange-quark couplings. Using instead the full likelihood~\cite{Gonzalez-Alonso:2016etj} given in \aref{LEFFE} slightly stronger constraints (and central values closer to zero) are obtained.} 
We stress however that the less sensitive experiments will be absolutely crucial to probe more independent directions in the space of dimension-6 operators. 
For the operators involving the 2nd generation lepton doublet the muon-neutrino scattering is a fairly sensitive probe. 
Again, $[O_{\ell q}^{(3)}]_{2211}$ is very precisely probed by the low-energy flavor observables because it affects the charged current. 
The sensitivity of low-energy experiments to the operators involving the right-handed muons is very poor. 
However, this is not a pressing problem, given these directions are very well probed by the LHC~\cite{deBlas:2013qqa}, as will be discussed in \sref{lhc}.
The $(ee)(qq)$ bounds shown in Table 5 are in excellent agreement with the 1-by-1 results of Ref.~\cite{deBlas:2013qqa}, whereas our $(\mu\mu)(qq)$ bounds are more stringent due to the inclusion of additional experimental input.

The LEP-2 constraints on operators involving 2nd generation or bottom quarks are similar as those shown in  \tref{1by1}.  We also give 1-by-1 constraints  on the chirality-violating LLQQ operators from the low-energy flavor observables: 
\beq
\bvec 
\, [c_{\ell e q u}]_{1111} \\ 
\, [c_{\ell e d q}]_{1111}  \\
\,[c^{(3)}_{\ell e q u}]_{1111}
\evec  = 
\bvec 
-\left ( 0.8  \pm 2.9 \right )\cdot 10^{-7} \\ 
\left ( 0.8  \pm 2.9 \right )\cdot 10^{-7} \\ 
\left ( 0.5  \pm 2.0 \right )\cdot 10^{-5}
\evec ,
\quad
\bvec 
\, [c_{\ell e q u}]_{2211} \\
\, [c_{\ell e d q}]_{2211}  \\ 
\, [c^{(3)}_{\ell e q u}]_{2211}
\evec  = 
\bvec 
\left ( 1.7  \pm 5.8 \right )\cdot 10^{-5} \\  
-\left ( 1.7   \pm 5.8 \right )\cdot 10^{-5} \\  
-\left ( 1.2  \pm 4.1 \right )\cdot 10^{-3} 
\evec .
\eeq
This exceptional sensitivity arises because these operators violate the approximate symmetries of the SM, leading  potentially to a large enhancement of several decays of low-mass hadrons.\footnote{%
More specifically they violate the approximate flavor symmetry of the SM $U(1)_\ell\times U(1)_e$ that suppresses the decay $\pi\to\ell\nu_\ell$ by a factor $m^2_{\ell}/\Lambda^2_{QCD}$. Thus, their bounds benefit from this large $\Lambda_{QCD}/m_{\ell}$ chiral enhancement. This does not apply however to the tensor operator $c^{(3)}_{\ell e q u}$, whose tree-level contribution to this specific decay is zero by simple Lorentz invariance considerations.
}
In particular, new physics generating the  pseudo-scalar $(ee)(qq)$ operator is probed up to $\Lambda/g_* \sim$100 TeV. 
Let us note that they dominate the $c^{(3)}_{\ell e q u}$ bounds shown above, despite the fact that they probe them only via 1-loop QED mixing~\cite{Voloshin:1992sn,Gonzalez-Alonso:2017iyc}.
For consistency with the rest of this work, these individual limits are obtained using $V=1$ at order $\Lambda^{-2}$. Working instead with the full  non-diagonal CKM matrix the limits are slightly modified, but more importantly one can set strong 1-by-1 limits in a long list of other (offdiagonal) operators. 

Finally, for the sake of completeness we show the 1-by-1 bound on the W coupling to right-handed 1st-generation quarks
\beq
\label{eq:GF_dgrwq1}
\, \delta g_R^{Wq_1} = -\left ( 3.9 \pm 2.9 \right )\cdot 10^{-4} ,
\eeq 
which is completely dominated by its contribution to the CKM-unitarity test of \eref{deltackm}.

%---------------------------------------------
\subsection{All out}
\label{sec:ao}

We now combine all the experimental observables summarized in \tref{inputsummary} along with the pole observables discussed in Ref.~\cite{Efrati:2015eaa}, which represent a total of 264 experimental input. 
These provide simultaneous constraints on 61 combinations of Wilson coefficients  of dimension-6 operators in the SMEFT Lagrangian (21 vertex corrections $\delta g$, 25  LLQQ and 15 LLLL operators) and on the $\tilde V_{ud}$ SM parameter. 
Marginalizing over $\tilde V_{ud}$ we find the following constraints:
\beq
\label{eq:AO}
\left(
\begin{array}{rcl}
\delta g^{We}_L \\ 
\delta g^{W\mu}_L \\ 
\delta g^{W\tau}_L \\ 
\delta g^{Ze}_L \\ 
\delta g^{Z\mu}_L \\ 
\delta g^{Z\tau}_L \\ 
\delta g^{Ze}_R  \\ 
\delta g^{Z\mu}_R \\ 
\delta g^{Z\tau}_R \\ 
\delta g^{Zu}_L \\ 
\delta g^{Zc}_L \\ 
\delta g^{Zt}_L \\ 
\delta g^{Zu}_R  \\ 
\delta g^{Zc}_R \\ 
\delta g^{Zd}_L \\ 
\delta g^{Zs}_L \\ 
\delta g^{Zb}_L \\
\delta g^{Zd}_R \\ 
\delta g^{Zs}_R \\ 
\delta g^{Zb}_R \\
\delta g^{Wq_1}_R  \\
 \, [c_{\ell \ell}]_{1111} \\ 
 \, [c_{\ell e}]_{1111} \\     
\,  [c_{e e}]_{1111}   \\    
\, [c_{\ell \ell}]_{1221} \\ 
\, [c_{\ell \ell}]_{1122} \\ 
 \, [c_{\ell e}]_{1122} \\   
 \, [c_{\ell e}]_{2211} \\    
\,  [c_{e e}]_{1122}  \\     
\, [c_{\ell \ell}]_{1331} \\ 
\, [c_{\ell \ell}]_{1133} \\ 
 \, [c_{\ell e}]_{1133} \\
 \,  [c_{\ell e}]_{3311} \\   
\,  [c_{e e}]_{1133}     \\ 
\, [\hat c_{\ell \ell}]_{2222} \\ 
\, [c_{\ell \ell}]_{2332}
\end{array}
\right) 
= 
\left(
\begin{array}{rcl}
\vphantom{\delta g^{We}_L} -1.00\pm 0.64 \\
\vphantom{\delta g^{W\mu}_L} -1.36\pm 0.59 \\
\vphantom{\delta g^{W\tau}_L} 1.95\pm 0.79 \\
\vphantom{\delta g^{Ze}_L} -0.023\pm 0.028 \\
\vphantom{\delta g^{Z\mu}_L} 0.01\pm 0.12\\
\vphantom{\delta g^{Z\tau}_L} 0.018\pm 0.059 \\
\vphantom{\delta g^{Ze}_R } -0.033\pm 0.027 \\
\vphantom{\delta g^{Z\mu}_R} 0.00\pm 0.14 \\
\vphantom{\delta g^{Z\tau}_R} 0.042\pm 0.062 \\
\vphantom{\delta g^{Zu}_L } -0.8\pm 3.1 \\
\vphantom{\delta g^{Zc}_L} -0.15\pm 0.36 \\
\vphantom{\delta g^{Zt}_L} -0.3\pm 3.8 \\
\vphantom{\delta g^{Zu}_R } 1.4\pm 5.1 \\
\vphantom{\delta g^{Zc}_R } -0.35\pm 0.53 \\
\vphantom{\delta g^{Zd}_L } -0.9\pm 4.4 \\
\vphantom{\delta g^{Zs}_L} 0.9\pm 2.8 \\
\vphantom{\delta g^{Zb}_L} 0.33\pm 0.17 \\
\vphantom{\delta g^{Zd}_R} 3\pm 16 \\
\vphantom{\delta g^{Zs}_R } 3.4\pm 4.9 \\
\vphantom{\delta g^{Zb}_R} 2.30\pm 0.88 \\
\vphantom{\delta g^{Wq_1}_R} -1.3\pm 1.7 \\
\vphantom{[c_{\ell \ell}]_{1111}} 1.01\pm 0.38 \\
\vphantom{[c_{\ell e}]_{1111}} -0.22\pm 0.22\\
\vphantom{[c_{e e}]_{1111}} 0.20\pm 0.38 \\
\vphantom{[c_{\ell \ell}]_{1221}} -4.8\pm 1.6 \\
\vphantom{[c_{\ell \ell}]_{1122}} 1.5\pm 2.1 \\
\vphantom{[c_{\ell e}]_{1122}} 1.5\pm 2.2 \\
\vphantom{[c_{\ell e}]_{2211}} -1.4\pm 2.2 \\
\vphantom{[c_{e e}]_{1122} } 3.4\pm 2.6 \\
\vphantom{[c_{\ell \ell}]_{1331}} 1.5\pm 1.3 \\
\vphantom{[c_{\ell \ell}]_{1133}} 0 \pm 11 \\
\vphantom{[c_{\ell e}]_{1133}} -2.3\pm 7.2 \\
\vphantom{[c_{\ell e}]_{3311} } 1.7\pm 7.2 \\
\vphantom{[c_{\ell \ell}]_{2332}} -1\pm 12 \\
\vphantom{[\hat c_{\ell \ell}]_{2222}} - 2 \pm 21 \\ 
\vphantom{[c_{\ell \ell}]_{2332}} 3.0\pm 2.3 
\end{array}
\right) \times 10^{-2} , 
\quad
\left(
\begin{array}{ccc}
 [c_{\ell q}^{(3)}]_{1111} \\  \, [\hat c_{e q}]_{1111} \\
   \, [\hat c_{\ell u}]_{1111} \\   \,   [\hat c_{\ell d}]_{1111}  \\   \,   [\hat c_{eu}]_{1111}  \\   \,  [\hat c_{ed}]_{1111}
    \\ 
  \,  [\hat c_{\ell q}^{(3)}]_{1122} \\   \,  [c_{\ell u}]_{1122}  \\   \,   [\hat c_{\ell d}]_{1122} \\  
  \,    [c_{eq}]_{1122}  \\   \,  [c_{eu}]_{1122} \\   \,  [\hat c_{ed}]_{1122} \\ 
 \,  [\hat c_{\ell q}^{(3)}]_{1133}  \\    \,  [c_{\ell d}]_{1133} \\   \,  [c_{eq}]_{1133} \\  \,   [c_{ed}]_{1133}
 \\ 
 \, [c_{\ell q}^{(3)}]_{2211} \\   \, [c_{\ell q}]_{2211} \\  \, [c_{\ell u}]_{2211} \\   \,   [c_{\ell d}]_{2211} 
 \\ \, [\hat c_{eq}]_{2211} 
 \\
 \, [c_{\ell e q u}]_{1111} \\ 
\, [c_{\ell e d q}]_{1111}  \\
\,[c^{(3)}_{\ell e q u}]_{1111}  \\ 
\, \eps_P^{d \mu}(2~\rm{GeV})
\end{array}
\right)  
= 
\left(
\begin{array}{ccc}
\vphantom{[c_{\ell q}^{(3)}]_{1111}} -2.2\pm 3.2 \\
\vphantom{[\hat c_{e q}]_{1111}} 100\pm 180 \\
\vphantom{[\hat c_{\ell u}]_{1111}} -5\pm 11 \\
\vphantom{[\hat c_{\ell d}]_{1111}} -5\pm 23 \\
\vphantom{ [\hat c_{eu}]_{1111}} -1\pm 12  \\
\vphantom{[\hat c_{ed}]_{1111}} -4\pm 21 \\
\vphantom{[\hat c_{\ell q}^{(3)}]_{1122}} -61\pm 32 \\
\vphantom{[c_{\ell u}]_{1122}} 2.4\pm 8.0 \\
\vphantom{[\hat c_{\ell d}]_{1122}} -310\pm 130  \\
\vphantom{[c_{eq}]_{1122} } -21\pm 28 \\
\vphantom{[c_{eu}]_{1122}} -87\pm 46 \\
\vphantom{[\hat c_{ed}]_{1122}} 270\pm 140 \\
\vphantom{ [\hat c_{\ell q}^{(3)}]_{1133}} -8.6\pm 8.0 \\
\vphantom{[c_{\ell d}]_{1133}} -1.4\pm 10 \\
\vphantom{[c_{eq}]_{1133}} -3.2\pm 5.1 \\
\vphantom{[c_{ed}]_{1133}} 18 \pm  20 \\
\vphantom{[c_{\ell q}^{(3)}]_{2211}} -1.2\pm 3.9 \\
\vphantom{[c_{\ell q}]_{2211}} 1.3\pm 7.6 \\
\vphantom{[c_{\ell u}]_{2211}} 15\pm 12 \\
\vphantom{[c_{\ell d}]_{2211}} 25\pm 34  \\
\vphantom{[\hat c_{eq}]_{2211} } 4 \pm 41  \\
\vphantom{[c_{\ell e q u}]_{1111}}-0.080\pm 0.075 \\
\vphantom{[c_{\ell e d q}]_{1111}} -0.079  \pm   0.074 \\
\vphantom{[c^{(3)}_{\ell e q u}]_{1111} }-0.02\pm 0.19 \\
\vphantom{\eps_P^{d \mu}(2~\rm{GeV})} -0.02\pm 0.15 \\
\end{array}
\right) \times 10^{-2}. 
\eeq 
The correlation matrix is available in the {\tt Mathematica} notebook attached as a supplemental material~\cite{magicnotebook}.
The complete Gaussian likelihood for the Wilson coefficients of dimension-6 SMEFT operators at the scale $\mu=m_Z$ can be reproduced from \eref{AO} and that correlation matrix. 
For user's convenience,  in the notebook the likelihood is displayed ready-made for cut and paste, and we also provide a translation to the Warsaw basis.   
That likelihood is relevant to constrain the masses and couplings of any new physics model whose leading effects at the weak scale can be approximated by tree-level contributions of vertex corrections and LLQQ and LLLL operators in the SMEFT.

The model-independent bounds on the vertex corrections are practically the same as the ones obtained from the pole observables only in Ref.~\cite{Efrati:2015eaa}. 
This is due to the fact that there are more 4-fermion operators than independent off-pole observables. 
Hence the latter serve to bound 4-fermion Wilson coefficients but cannot further constrain $\delta g$.
Nevertheless, there are non-zero correlations between the constraints on vertex corrections and 4-fermion operators that are captured by our analysis. 
It is worth stressing the CKM-unitarity test $\Delta_{CKM}$ of \eref{deltackm}, which actually provides stronger one-by-one limits on the vertex corrections $\delta g_L^{W q_1}$ and $\delta g_L^{W \mu}$ than all pole observables combined.

Furthermore, the low-energy flavor observables provide a percent level bound on the W boson coupling to right-handed light quarks $\delta g^{W q_1}_R$~\cite{Gonzalez-Alonso:2016etj}. 
Recall that $\delta g^{W q}_R$ are not probed by the pole observables at tree level and $\cO(\Lambda^{-2})$ in the SMEFT expansion, therefore the model-independent limit in \eref{AO} (from Ref.~\cite{Gonzalez-Alonso:2016etj}) is a new result.  
It is weaker than the one in \eref{GF_dgrwq1} because in the global fit the strong constraints from the CKM-unitarity test of \eref{deltackm} are diluted by marginalizing over less precisely probed dimension-6 parameters.  
Nevertheless, the constraint on $\delta g^{W q_1}_R$ will typically be stronger in specific new physics scenarios, unless they predict that the particular linear combination on the r.h.s of \eref{deltackm} approximately vanishes at the sub-per-mille level.   
     
The bounds on LLLL operators involving only electrons and/or muons  are also similar  to the ones previously obtained in Ref.~\cite{Falkowski:2015krw}, with the exception of $[c_{\ell\ell}]_{2222}$ which is now bound due to the inclusion of neutrino trident production data. 
For the $ee\tau\tau$ operators the bounds are much stronger thanks to including the VENUS $\tau \tau$ polarization data, which resolves the degeneracies  present in the fit  of Ref.~\cite{Falkowski:2015krw}.

The model-independent bounds on LLQQ operators in  \eref{AO} are new. 
Previous global SMEFT analyses targeting these operators \cite{Han:2004az,Han:2005pr,Berthier:2015gja} were carried out assuming some simplifying flavor structure, such as the $U(3)^5$ symmetry~\cite{Han:2004az}, which greatly reduces the number of independent Wilson coefficients. On the other hand, previous analyses working in a flavor general setup provided 1-by-1 bounds (see e.g. Ref.~\cite{Carpentier:2010ue,deBlas:2013qqa}). Thus, the global bounds applicable for a completely arbitrary flavor structure are obtained for the first time in this paper, and they represent our main result. 
They are relevant for a large class of new physics scenarios with or without approximate flavor symmetries.   
In particular, models addressing various flavor anomalies necessarily do not respect the $U(3)^5$ symmetry, and therefore the global likelihood we obtained may provide new constraints on their parameters.   

We find several poorly constrained directions in the space of  LLQQ operators. 
As discussed earlier, $ [\hat c_{e q}]_{1111}$ is currently  constrained only by very imprecise measurements of electron neutrino scattering on nucleons, such that  the experiments are insensitive to $ [\hat c_{e q}]_{1111} \lesssim 1$. 
More surprisingly, another practically unconstrained direction emerges in our fit,  which  roughly corresponds to the linear combination
$[\hat c_{ed} + 0.6\,\hat c_{\ell d}]_{1122}$.
This can be traced to the fact that the LEP-2 collider was scanning a fairly narrow range of $\sqrt{s}$ in $e^+ e^-$ collisions.  
For this reason, not all theoretically available combinations discussed in \sref{GF_flat} are resolved in practice.
Again, it is should be noted that constraints in typical scenarios generating these LLQQ operators  will be stronger unless the operators accidentally align with the flat directions in our  fit. 
We stress that the global likelihood provided in the supplemental material  \cite{magicnotebook} retains the full information about the correlations.

%%%%%%%%%%%%%%%%%%%%%%
\subsection{Flavor-universal limit}

The general likelihood presented  in \sref{ao} can be easily  restricted to a smaller subspace relevant for any particular scenario. 
We present here the results for the flavor-universal limit, where dimension-6 operators are invariant under the global flavor symmetry $U(3)^5$. 
The symmetry implies that 1) all off-diagonal and chirality-violating operators as well as $\delta g^{Wq}_R$ are absent,  2) the remaining operators do not  carry the flavor index. 
The only subtlety concerns the $[c_{\ell \ell}]_{IJKL}$ coefficients, since two independent contractions of flavor indices are allowed by the $U(3)^5$ symmetry. We follow the common practice of parametrizing them in terms of the two $U(3)^5$-symmetric operators $O_{\ell\ell} \equiv \frac{1}{2} \sum_{I,J} (\bar \ell_I \bar\sigma_\mu \ell_I) (\bar \ell_J \bar\sigma_\mu \ell_J)$ and $O^{(3)}_{\ell\ell} \equiv \frac{1}{2} \sum_{I,J} (\bar \ell_I \sigma^i \bar\sigma_\mu \ell_I) (\bar \ell_J \sigma^i \bar\sigma_\mu \ell_J)$. 
All in all, with the parameterization of the dimension-6 space used in this paper, the $U(3)^5$ symmetry corresponds to the following pattern: 
\beq
\label{eq:u5params}
\bvec 
\delta g^{We_J}_L \\ 
\delta g^{Ze_J}_L \\ 
\delta g^{Ze_J}_R  \\ 
\delta g^{Zu_J}_L \\ 
\delta g^{Zu_J}_R  \\ 
\delta g^{Zd_J}_L \\ 
\delta g^{Zd_J}_R 
\evec
= 
\bvec 
\vphantom{g^{We_J}_L} \delta g^{We}_L \\ 
\vphantom{g^{We_J}_L} \delta g^{Ze}_L \\ 
\vphantom{g^{We_J}_L} \delta g^{Ze}_R  \\ 
\vphantom{g^{We_J}_L} \delta g^{Zu}_L \\ 
\vphantom{g^{We_J}_L} \delta g^{Zu}_R  \\ 
\vphantom{g^{We_J}_L} \delta g^{Zd}_L \\ 
\vphantom{g^{We_J}_L} \delta g^{Zd}_R 
\evec
, \, 
\bvec
\vphantom{c_{\ell \ell}^{(3)}} \, [c_{\ell \ell}]_{JJJJ} \\ 
\vphantom{c_{\ell \ell}^{(3)}}  \, [c_{\ell \ell}]_{IJJI} \\ 
\vphantom{c_{\ell \ell}^{(3)}}   \, [c_{\ell \ell}]_{IIJJ} \\ 
 \, [c_{\ell e}]_{IIJJ} \\     
\,  [c_{e e}]_{IIJJ}    
\evec
=  
\bvec
 c_{\ell \ell}  +  c_{\ell \ell}^{(3)}  \\
 2  c_{\ell \ell}^{(3)} \\ 
c_{\ell \ell}  -c_{\ell \ell}^{(3)}   \\
 c_{\ell e} \\     
c_{e e} 
\evec
, \, 
\bvec
\, [c_{\ell q}^{(3)}]_{IIJJ} \\  \, [c_{\ell q}]_{IIJJ} \\   \, [c_{e q}]_{IIJJ} \\
   \, [c_{\ell u}]_{IIJJ} \\   \,   [c_{\ell d}]_{IIJJ}  \\   \,   [c_{eu}]_{IIJJ}  \\   \,  [c_{ed}]_{IIJJ}
\evec 
=
\bvec
c_{\ell q}^{(3)} \\  c_{\ell q} \\   c_{e q} \\
   c_{\ell u} \\   c_{\ell d}  \\   c_{eu}  \\   c_{ed}
\evec , 
\eeq 
and all the remaining vertex corrections and 4-fermion Wilson coefficients vanish. 
This setup corresponds to the SMEFT limit studied in the pioneering work of Ref.~\cite{Han:2004az}.\footnote{Let us note that the more recent analysis of Ref.~\cite{Berthier:2015gja} corresponds to a more restricted scenario, since the two independent coefficients $c_{\ell \ell}$ and $c_{\ell \ell}^{(3)}$ are controlled by one single coefficient $C_{\ell\ell}$ in that work.}

It turns out that the global likelihood constrains the entire restricted parameter set introduced in \eref{u5params}. 
Thus, unlike in the flavor-generic case,  there is no need to define new variables $\hat c$ in order to factor out the flat directions.  
Marginalizing over $\tilde V_{ud}$, we find  the following constraints:  
 \beq
 \label{eq:dgu5}
 \bvec 
\delta g^{We}_L \\ 
\delta g^{Ze}_L \\ 
\delta g^{Ze}_R  \\ 
\delta g^{Zu}_L \\ 
\delta g^{Zu}_R  \\ 
\delta g^{Zd}_L \\ 
\delta g^{Zd}_R 
\evec 
= \left(
\begin{array}{ccc}
 -1.22 & \pm  & 0.81 \\
 -0.10 & \pm  & 0.21 \\
 -0.15 & \pm  & 0.23 \\
 -1.6 & \pm  & 2.0 \\
 -2.1 & \pm  & 4.1 \\
 1.9 & \pm  & 1.4 \\
 15 & \pm  & 7 \\
\end{array}
\right) \times 10^{-3},
 \eeq
 \beq
  \label{eq:cu5}
 \bvec
c_{\ell \ell}^{(3)} \\ 
 c_{\ell \ell}   \\ 
 c_{\ell e} \\     
c_{e e} 
\evec
=\left(
\begin{array}{ccc}
 -3.0 & \pm  & 1.7 \\
 7.2 & \pm  & 3.3 \\
 0.2 & \pm  & 1.3 \\
 -2.5 & \pm  & 3.0 \\
\end{array}
\right) \times 10^{-3}, 
\quad
\bvec
c_{\ell q}^{(3)} \\  c_{\ell q} \\   c_{e q} \\
   c_{\ell u} \\   c_{\ell d}  \\   c_{eu}  \\   c_{ed}
\evec  = 
\left(
\begin{array}{ccc}
 -4.8 & \pm  & 2.3 \\
 -15.4 & \pm  & 9.1 \\
 -14 & \pm  & 23 \\
 4 & \pm  & 24 \\
 6 & \pm  & 42  \\
 4 & \pm  & 11  \\
 26 & \pm  & 18 \\
\end{array}
\right) \times 10^{-3}. 
 \eeq
The  correlation matrix reads $\rho$ = 
 \beq 
 \label{eq:rhou5}
 \scriptsize
 \left(
\begin{array}{cccccccccccccccccc}
 1. & -0.5 & 0.2 & 0.1 & 0.1 & 0. & 0. & 1. & -0.5 & 0. & -0.1 & 0.4 & -0.1 & 0. & 0.1 & 0. & 0.1 & 0. \\
 -0.5 & 1. & 0.3 & -0.1 & 0. & -0.2 & -0.2 & -0.5 & 0.2 & 0. & 0.1 & -0.1 & 0.1 & 0. & 0. & 0. & -0.1 & -0.1 \\
 0.2 & 0.3 & 1. & 0. & 0. & -0.3 & -0.3 & 0.2 & -0.2 & 0. & 0.1 & 0.3 & 0. & 0.1 & 0.1 & 0.1 & 0. & -0.1 \\
 0.1 & -0.1 & 0. & 1. & 0.8 & 0.2 & 0.1 & 0.1 & 0. & 0. & 0. & 0.7 & -0.3 & 0. & 0.1 & 0. & 0.5 & 0.1 \\
 0.1 & 0. & 0. & 0.8 & 1. & 0.1 & 0.2 & 0.1 & 0. & 0. & 0. & 0.7 & -0.3 & 0. & 0.1 & 0. & 0.5 & 0.2 \\
 0. & -0.2 & -0.3 & 0.2 & 0.1 & 1. & 0.9 & 0. & 0. & 0. & 0. & -0.4 & -0.2 & -0.1 & -0.1 & -0.2 & 0.2 & 0.4 \\
 0. & -0.2 & -0.3 & 0.1 & 0.2 & 0.9 & 1. & 0. & 0. & 0. & 0. & -0.5 & -0.2 & -0.1 & -0.1 & -0.2 & 0.2 & 0.4 \\
 1. & -0.5 & 0.2 & 0.1 & 0.1 & 0. & 0. & 1. & -0.5 & 0. & -0.1 & 0.4 & -0.1 & 0. & 0.1 & 0. & 0.1 & 0. \\
 -0.5 & 0.2 & -0.2 & 0. & 0. & 0. & 0. & -0.5 & 1. & -0.2 & -0.6 & -0.2 & 0. & 0. & 0. & 0. & 0. & 0. \\
 0. & 0. & 0. & 0. & 0. & 0. & 0. & 0. & -0.2 & 1. & -0.2 & 0. & 0. & 0. & 0. & 0. & 0. & 0. \\
 -0.1 & 0.1 & 0.1 & 0. & 0. & 0. & 0. & -0.1 & -0.6 & -0.2 & 1. & 0. & 0. & 0. & 0. & 0. & 0. & 0. \\
 0.4 & -0.1 & 0.3 & 0.7 & 0.7 & -0.4 & -0.5 & 0.4 & -0.2 & 0. & 0. & 1. & -0.1 & 0.1 & 0.2 & 0.1 & 0.3 & -0.1 \\
 -0.1 & 0.1 & 0. & -0.3 & -0.3 & -0.2 & -0.2 & -0.1 & 0. & 0. & 0. & -0.1 & 1. & -0.2 & -0.7 & -0.6 & -0.5 & -0.9 \\
 0. & 0. & 0.1 & 0. & 0. & -0.1 & -0.1 & 0. & 0. & 0. & 0. & 0.1 & -0.2 & 1. & 0.7 & 0.9 & -0.5 & 0.5 \\
 0.1 & 0. & 0.1 & 0.1 & 0.1 & -0.1 & -0.1 & 0.1 & 0. & 0. & 0. & 0.2 & -0.7 & 0.7 & 1. & 0.9 & -0.1 & 0.8 \\
 0. & 0. & 0.1 & 0. & 0. & -0.2 & -0.2 & 0. & 0. & 0. & 0. & 0.1 & -0.6 & 0.9 & 0.9 & 1. & -0.2 & 0.7 \\
 0.1 & -0.1 & 0. & 0.5 & 0.5 & 0.2 & 0.2 & 0.1 & 0. & 0. & 0. & 0.3 & -0.5 & -0.5 & -0.1 & -0.2 & 1. & 0.3 \\
 0. & -0.1 & -0.1 & 0.1 & 0.2 & 0.4 & 0.4 & 0. & 0. & 0. & 0. & -0.1 & -0.9 & 0.5 & 0.8 & 0.7 & 0.3 & 1. \\
\end{array}
\right)
 \eeq 
 where the rows and columns correspond to the ordering of the parameters in \eref{dgu5} and \eref{cu5}.  
The correlation matrix with more significant digits (necessary for practical applications) is given in the {\tt Mathematica} notebook attached as supplemental material  \cite{magicnotebook}.

Thanks to lifting the exact and approximate flat directions, in the $U(3)^5$ symmetric limit typical constraints on the dimension-6 parameters are at the per-mille level. 
We note that  the vertex corrections are constrained slightly better than when only the pole observables are used \cite{Efrati:2015eaa}, thanks to the precise input from low-energy flavor measurements. Most of the LLQQ operators are constrained at the percent level. 
 
Also working in the flavor-universal limit, Ref.~\cite{Falkowski:2015jaa} obtained bounds on 10 \textit{additional} SMEFT coefficients using Higgs data and $WW$ production at LEP2. The only flavor-universal SMEFT coefficients unconstrained by these two fits are those that are either CP-violating, or contain only quarks, only gluons or only higgses.

 %%%%%%%%%%%%%%%%%%%%%%
\subsection{Oblique parameters} 

In the literature, precision constraints on new physics are often quoted in the language of {\em oblique parameters} $S$, $T$, $W$, $Y$ \cite{Peskin:1991sw,Barbieri:2004qk}.
These correspond to a further restriction of the pattern of the dimension-6 parameters in the $U(3)^5$ symmetric case \cite{Wells:2015uba,Falkowski:2015krw}: 
 \bea
\label{eq:UNI_STWYtoHB}
\delta g^{Z f}_{L/R} &=&   \alpha  \left \{ 
T^3_{f_{L/R}} { T - W - {g_Y^2\over g_L^2}  Y \over 2}
+ Q_f { 
 2 g_Y^2  T  - (g_L^2 + g_Y^2)  S  + 2 g_Y^2 W + {2 g_Y^2 (2 g_L^2 - g_Y^2) \over g_L^2}  Y   \over 4 (g_L^2 - g_Y^2)} \right \}, 
 \nnl 
\delta g^{W  e}_L &=& {\alpha \over 2 (g_L^2 - g_Y^2)} \left (
- {g_L^2 + g_Y^2 \over 2} S  + g_L^2 T -(g_L^2 - 2 g_Y^2) W + g_Y^2 Y \right ), 
 \nnl 
c_{\ell \ell}^{(3)}  &=&  c_{\ell q}^{(3)}  =c_{q q}^{(3)}  =   -  \alpha  W,  
\qquad c_{f_1 f_2}  =    - 4 Y_{f_1} Y_{f_2} {g_Y^2\over g_L^2}   \alpha  Y ~,
\eea 
where $Y_{f_i}$ is the fermionic hypercharge. 
With this pattern, all vertex corrections and 4-fermion operators can be redefined away, such that new physics affects only the electroweak gauge boson propagators. 
Restricting the $U(3)^5$ symmetric likelihood using \eref{UNI_STWYtoHB}  we find the following constraints on the oblique parameters:
\beq
\label{eq:styw}
\bvec
S \\ T \\ Y \\ W 
\evec 
= \bvec -0.10 \pm 0.13 \\ 0.02 \pm 0.08 \\ -0.15 \pm 0.11 \\ -0.01 \pm 0.08 \evec, 
\qquad 
\rho = \left(
\begin{array}{cccc}
 1. & 0.86 & 0.70 & -0.12 \\
 . & 1. & 0.39 & -0.06 \\
 . & . & 1. & -0.49 \\
 . & . & . & 1. \\
\end{array}
\right). 
\eeq 
The constraints on the oblique corrections are dominated by the pole-observables and lepton-pair production in LEP-2.  
The new observables probing LLQQ operators  do not affect these constraints significantly. 
In particular, the low-energy flavor observables do not probe the oblique corrections at all. 
Compared to the fit in Ref.~\cite{Falkowski:2015krw}, we only observe a small shift of the central values.\footnote{%  
The $\cO(10\%)$ increase of some errors compared to \cite{Falkowski:2015krw} is due to using different values of the electroweak couplings to evaluate the dimension-6 shifts of the LEP-2 observables. }

%%%%%%%%%%%%%%%%%%%%%%%%%%%%%%%
\section{Comments on LHC reach}
\label{sec:lhc}

Four-fermion LLQQ operators can be probed via the $q \bar q \to \ell^+ \ell^-$ processes in hadron colliders. 
Previously several groups set bounds on their Wilson coefficients through the reanalysis within the SMEFT of various ATLAS and CMS exotic searches (see e.g.~\cite{Cirigliano:2012ab,deBlas:2013qqa,Greljo:2017vvb}). 
%It is not our goal here to improve on these analyses. We only present a brief comparison between the sensitivity of the LHC run-1 and of the low-energy observables discussed in this paper
In this section we  derive analogue bounds using the recently published measurements of the differential Drell-Yan cross sections in the dielectron and dimuon channels~\cite{Aad:2016zzw}. Our main goal here is to present a brief comparison between the sensitivity of the LHC run-1 and of the low-energy observables discussed in this paper.

Precision measurements in hadron collider environments are challenging. 
Individual observables are typically measured with much worse accuracy than in lepton colliders or very low-energy experiments.   
However, the effect of 4-fermion operators on scattering amplitudes grows with the collision energy $E$ as 
$\sim c_{4f} E^2/v^2$. 
As a consequence, the superior  energy reach of the LHC  compensates the inferior precision in this case~\cite{Cirigliano:2012ab,deBlas:2013qqa}. 
This message  was recently stressed in Ref.~\cite{Farina:2016rws} in the context of the determination of the oblique parameters, which encode new physics corrections to propagators of the electroweak gauge bosons.    
It turns out that the effect of the oblique parameters $W$ and $Y$ \cite{Barbieri:2004qk} on the high invariant-mass  tail of  ${d \sigma (pp \to \ell^+ \ell^-) \over d m_{\ell \ell}}$ also grows with $E$ (as opposed to that of the more familiar $S$ and $T$ parameters \cite{Peskin:1991sw}). 
The current LHC constraint on $W$ and $Y$ are already competitive with those obtained from low-energy precision experiments, and will become more accurate with the full run-2 dataset at $\sqrt{s} \approx 13$-$14$~TeV \cite{Farina:2016rws}.
In the SMEFT framework, $W$ and $Y$ correspond to a particular pattern of vertex corrections and 4-fermion operators \cite{Wells:2015uba,Falkowski:2015krw}, \textit{cf.} \eref{UNI_STWYtoHB}.
Therefore we expect that similar arguments apply, and that competitive bounds on the LLQQ operators can be extracted from ATLAS and CMS measurements of  ${d \sigma (pp \to \ell^+ \ell^-) \over d m_{\ell \ell}}$.  Below we present some quantitative illustrations of this message.

\begin{table}
\bc
$(ee)(qq)$ \\
\footnotesize
\begin{tabular}{|c|c|c|c|c|c|c|c|c}
\hline  
&  $ [c_{\ell q}^{(3)}]_{1111} $ &  $[c_{\ell q}]_{1111}$   & $[c_{\ell u}]_{1111}$  & $[c_{\ell d }]_{1111}$ 
& $[c_{eq}]_{1111}$  & $[c_{eu}]_{1111}$ &   $[c_{ed}]_{1111}$  \\ 
\hline 
Low-energy
&   $0.45 \pm 0.28$ &  $ 1.6 \pm 1.0 $ &  $ 2.8 \pm 2.1$ &  $3.6 \pm 2.0$ &  $ -1.8 \pm 1.1 $ &  $-4.0 \pm 2.0 $ &  $-2.7 \pm 2.0$ 
\\ \hline 
LHC${}_{1.5}$
& $-0.70^{+0.66}_{-0.74}$ & $2.5^{+1.9}_{-2.5}$  & $2.9^{+2.4}_{-2.9}$ &  $-1.6^{+3.4}_{-3.0}$ & $1.6^{+1.8}_{-2.2}$ &  $1.6^{+2.5}_{-1.5}$ &  $-3.1^{+3.6}_{-3.0}$   
 \\ \hline 
LHC${}_{1.0}$
& $-0.84^{+0.85}_{-0.92}$ & $3.6^{+3.6}_{-3.7}$  & $4.4^{+4.4}_{-4.7}$ &  $-2.4^{+4.8}_{-4.7}$ & $2.4^{+3.0}_{-3.2}$ &  $1.9^{+2.5}_{-1.9}$ &  $-4.6^{+5.4}_{-4.1}$   
 \\ \hline 
LHC${}_{0.7}$
& $-1.0^{+1.4}_{-1.5}$ & $5.9 \pm 7.2$  & $7.4\pm 9.0$ &  $-3.6\pm 8.7$ & $3.8 \pm 5.9$ &  $2.1^{+3.8}_{-2.9}$ &  $-8 \pm 10$   
 \\ \hline 
\end{tabular}
\ec
\bc 
$(\mu \mu)(qq)$ \\
\footnotesize
\begin{tabular}{|c|c|c|c|c|c|c|c|}
\hline
&  $ [c_{\ell q}^{(3)}]_{2211} $ &  $[c_{\ell q}]_{2211}$   & $[c_{\ell u}]_{2211}$  & $[c_{\ell d }]_{2211}$  
& $[c_{eq}]_{2211}$  & $[c_{eu}]_{2211}$ &   $[c_{ed}]_{2211}$\\   
\hline 
Low-energy
& $ -0.2 \pm 1.2$  & $4 \pm 21$ & $18 \pm 19$ & $-20 \pm 37$ &
 $40 \pm 390$ & $-20 \pm 190$ & $40 \pm 390$    \\
\hline
LHC${}_{1.5}$
& $-1.22^{+0.62}_{-0.70}$ &  $1.8 \pm 1.3$  & $2.0\pm 1.6$ &  $-1.1 \pm 2.0$ & $1.1 \pm 1.2$ &  $2.5^{+1.8}_{-1.4}$ &  $-2.2 \pm 2.0$  \\  
\hline
LHC${}_{1.0}$
& $-0.72^{+0.81}_{-0.87}$ &  $3.2^{+4.0}_{-3.5}$  & $3.9^{+4.8}_{-4.4}$ &  $-2.3^{+4.9}_{-4.7}$ & $2.3^{+3.1}_{-3.2}$ &  $1.6^{+2.3}_{-1.8}$ &  $-4.4 \pm 5.3$   
 \\ \hline 
LHC${}_{0.7}$
& $-0.7^{+1.3}_{-1.4}$ &  $3.2^{+10.3}_{-4.8}$  & $4.3^{+12.5}_{-6.4}$ &  $-3.6 \pm 9.0$ & $3.8 \pm 6.2$ &  $1.6^{+3.4}_{-2.7}$ &  $-8 \pm 11$   
 \\ \hline 
\end{tabular}
\ec 
\bc 
%\phantom{()}
Chirality-violating operators ($\mu=$ 1 TeV) \\
\footnotesize
\begin{tabular}{|c|c|c|c|c|c|c|}
\hline
& $[c_{\ell e q u}]_{1111} $ 
& $[c_{\ell e d q}]_{1111} $
& $[c^{(3)}_{\ell e q u}]_{1111} $
& $[c_{\ell e q u}]_{2211} $
& $[c_{\ell e d q}]_{2211} $
& $[c^{(3)}_{\ell e q u}]_{2211}$ \\   
\hline 
Low-energy &  $(-0.6 \pm 2.4)10^{-4}$   & $(0.6 \pm 2.4)10^{-4}$  & $(0.4 \pm 1.4)10^{-3}$   & $0.014(49)$   & $-0.014(49)$  & $-0.09(29)$
\\ \hline  
LHC${}_{1.5}$
& $0 \pm 2.0$ & $0 \pm 2.6$ &  $0\pm 0.91$ &  $0 \pm 1.2$  & $0 \pm 1.6$ &     $0 \pm 0.56$
 \\ \hline
LHC${}_{1.0}$
& $0 \pm 2.9$ & $0 \pm 3.7$ &  $0\pm 1.4$ &  $0 \pm 2.9$  & $0 \pm 3.7$ &     $0 \pm 1.4$
 \\ \hline 
LHC${}_{0.7}$
& $0 \pm 5.3$ & $0 \pm 6.6$ &  $0\pm 2.6$ &  $0 \pm 5.5$  & $0 \pm 6.9$ &     $0 \pm 2.6$
 \\ \hline 
\end{tabular}
\ec 
\caption{
Comparison of low-energy and LHC constraints (in units of $10^{-3}$) on the Wilson coefficients of the chirality-conserving $(ee)(qq)$ and $(\mu \mu)(qq)$  and chirality-violating  operators defined at the scale $\mu = 1$~TeV. 
The $68\%$~CL bounds are derived assuming only one 4-fermion operator is present at a time, and that the vertex corrections and $[c_{\ell \ell}]_{1221}$ are absent.
The low-energy constraints combine all experimental  input summarized in \tref{inputsummary}. 
The LHC${}_{1.5}$ constraints use the $m_{\ell \ell} \in [0.5$-$1.5]$~TeV bins of the measured differential $e^+ e^-$ and $\mu^+ \mu^-$ cross sections at the 8 TeV LHC \cite{Aad:2016zzw}. 
We also separately show the constraints obtained when the $m_{\ell \ell} \in [0.5$-$1.0]$~TeV  (LHC${}_{1.0}$) and $m_{\ell \ell} \in [0.5$-$0.7]$~TeV (LHC${}_{0.7}$) data range is used.  
%`x' signals that the operator is not probed at tree level by that (combination of) experiment(s).
}
\label{tab:lhc}
\end{table}

In the situation when only one LLQQ operator is present at a time and all other dimension-6 operators are absent, the sensitivity of the LHC run-1 and of the low-energy observables is contrasted in \tref{lhc}.
To estimate the LHC reach  we use 3 bins in the range  $m_{\ell \ell} \in [0.5$-$1.5]$~TeV of the ATLAS measurement of the differential $e^+ e^-$ and $\mu^+ \mu^-$ cross sections at the 8 TeV LHC (20.3 fb$^{-1}$) \cite{Aad:2016zzw}.  
This is shown under the label of LHC${}_{1.5}$ constraints in \tref{inputsummary},  and it is compared to the combined constraints using the low-energy input. 
For the chirality conserving $(ee)(qq)$ operators the two are indeed similarly sensitive.
For the chirality conserving $(\mu \mu)(qq)$ operators the low-energy bounds are relatively  weaker, especially  for the operators that do not affect the muon neutrino couplings. 
With the exception of $[O^{(3)}_{\ell q}]_{2211}$ probed by the flavor observables, the LHC sensitivity is superior by at least an order of magnitude. 
Therefore in these directions in the parameter space of dimension-6 SMEFT the LHC is in a completely uncharted territory.
The situation is quite opposite for the chirality-violating  $(ee)(qq)$ and $(\mu \mu)(qq)$  operators. 
There the light quark transitions offer a superior sensitivity with which the LHC cannot compete in most cases. 
The exception is the  $[O^{(3)}_{\ell e qu}]_{2211}$ operator where the LHC reach is comparable. 

An important difference between the LHC and low-energy constraints should be emphasized. 
The latter are obtained in the energy regime where it is very plausible to assume the validity of the EFT. 
Here, by validity we mean that the SMEFT with dimension-6 operators adequately  describes the physics of the underlying UV completion. 
First of all, if such completion contains new states at $\sim 1$ TeV then clearly the LHC bounds in \tref{lhc} cannot be applied and a model-dependent approach becomes necessary. This is however not the case for the SMEFT bounds derived from low-energy data in the previous section, which are still valid. 
On the other hand, even in the absence of such \lq\lq light\rq\rq~states one should analyze the sensitivity to $\cO(\Lambda^{-4})$ terms. 
The precisely measured low-energy observables are dominated by $\cO(\Lambda^{-2})$ contributions of dimension-6 operators, whereas the quadratic terms in the Wilson coefficients, formally $\cO(\Lambda^{-4})$, are negligible. 
In contrast, the one-by-one LHC constraints on 4-fermion operators in \tref{lhc} have in general a similar sensitivity to linear and quadratic terms.\footnote{
In fact, in a few LHC$_{0.7}$ entries in \tref{lhc} there is an additional (not shown) second solution far from the origin.}
Notice that this problem becomes much more severe in a global fit and that in the particular case of the chirality-violating operators there is no interference at all. 
This may undermine the SMEFT $1/\Lambda^2$ expansion for generic UV completions, and it is not clear whether the dimension-8 and higher operators can be neglected in the analysis.  
As discussed in Ref.~\cite{Contino:2016jqw}, in such a case the EFT is still valid for strongly coupled UV completions, where the dimension-6 squared terms are parametrically enhanced with respect to the dimension-8 contributions by a large new physics coupling.
On the other hand,  for weakly coupled UV completion one should use weaker LHC bounds obtained by truncating the $\sqrt{s}$ range of the analyzed data at some $M_{\rm cut}$ above which the SMEFT is no longer valid. 
For illustration,  in \tref{inputsummary} we show the analogous LHC constraints with $M_{\rm cut} = 1$~TeV (LHC${}_{1.0}$) and $M_{\rm cut} = 0.7$~TeV (LHC${}_{0.7}$). 

Another practical consequence of the quadratic terms domination at the LHC is that the likelihood for the Wilson coefficients is not approximately Gaussian.
That means it is not fully characterized by the central values,  1~$\sigma$ errors, and the correlation matrix, 
as is the case for the low-energy observables.  
This  makes the presentation of the global fit results more cumbersome. 

Last, let us notice that the dilepton-production cross section is also sensitive to SMEFT coefficients that are flavor non-diagonal in the quark bilinear if we go beyond the $V=1$ approximation at order $\Lambda^{-2}$. This was exploited in Ref.~\cite{Gonzalez-Alonso:2016etj} to set bounds on the Wilson coefficients of chirality-violating $\ell\ell21$ operators.

%%%%%%%%%%%%%%%%%%%%%%%%%%%%%%%
\section{Conclusions}
\label{sec:concl}
 
This paper compiles information from a number of experiments sensitive to flavor-conserving LLQQ operators. 
The main focus is on experiments probing physics well below the weak scale, such as 
neutrino scattering on  nucleon targets, atomic parity violation, parity-violating electron scattering on nuclei, 
and so on.  
Information from $e^+ e^-$ collisions at the center-of-mass energies around the weak scale is also included. 
This is combined with previous analyses studying 4-lepton operators and the strength of the Z and W boson couplings to matter.  
The ensemble of data is interpreted as constraints on heavy  new physics encoded in tree-level effects of dimension-6 operators in the SMEFT. 
The main strength of this analysis is that we allow all independent operators to be simultaneously present with an arbitrary flavor structure. 
Another novelty is the inclusion of low-energy flavor constraints from pion, neutron, and nuclear decays, 
recently summarized in Ref.~\cite{Gonzalez-Alonso:2016etj}.
The leading renormalization group running effects from low energies to the weak scale are taken into account.  

We obtain simultaneous constraints on 61 linear combinations of Wilson coefficients in the SMEFT. 
The results are presented as a multi-dimensional likelihood function, which is provided in a {\tt Mathematica} notebook attached as supplemental material  \cite{magicnotebook}.  
The likelihood can easily be projected onto more restricted new physics scenarios. 
As an illustration, we provide constraints on the SMEFT operators in the $U(3)^5$-symmetric scenario, 
and on the oblique parameters $S$, $T$, $W$, $Y$. 
The likelihood  can be  used to place limits on masses and couplings in a large class of theories beyond the SM when  the mapping between these theories and the SMEFT is known. 

Finally, a brief comparison of the sensitivity  of low-energy experiments to  LLQQ operators with that of the LHC is provided. 
For many directions in the SMEFT parameters space, dilepton production at the LHC  is exploring virgin territories not constrained by previous experiments.
This is especially true for the chirality-conserving $2 \mu 2 q$ operators, where $q$ are light quarks, while for the chirality-conserving $2 e 2 q$ operators the LHC and low-energy probes are similarly sensitive. 
It would be beneficial to recast the LHC dilepton results in a model-independent form of a global likelihood on the SMEFT Wilson coefficients. 
We leave this task for future publications.  

The SMEFT constraints summarized in this paper should be improved in the near future.
Measurements of the differential Drell-Yan production cross sections at the LHC run-2 will provide a more powerful probe of LLQQ operators, thanks to the increased center-of-mass  energy of the collisions and higher luminosities.\footnote{As the recent recast of 13-TeV ATLAS data carried out in Ref.~\cite{Greljo:2017vvb} shows, this is already the case with the currently available luminosity (36.1 fb$^{-1}$). Expected bounds with 3000 fb$^{-1}$ of data can also be found in that work.} 
Progress is imminent on the low-energy front as well,  e.g. thanks to more precise measurements  of low-energy electron scattering in the Q-weak, MOLLER and P2 experiments.  
In this paper we have stressed the importance of  probing new physics in multiple low- and high-energy experiments.  
The huge number of independent SMEFT operators requires a rich and diverse set of  observables in order to lift flat directions in the global likelihood.
In fact, several poorly or not-at-all constrained directions in the SMEFT parameter space persist, 
as is evident from \eref{AO}. 
This is especially true for operators involving the second and third generation quarks or the third generation leptons, but some flat directions involve the first generation fermions.   
The existence of these unexplored directions could be an inspiration to design new experiments and observables.

\appendix 

\renewcommand{\theequation}{\Alph{section}.\arabic{equation}}

%%%%%%%%%%%%%%%%%%%%%%%%%%%%
%\section{Likelihood, correlations, eigenvalues, eigenvectors}
%\label{app:likelihood}
%\setcounter{equation}{0}  

%%%%%%%%%%%%%%%%%%%%%%%%%%%%
\section{Translation to Warsaw Basis}
\label{app:warsaw}
\setcounter{equation}{0}  

In this paper we parametrize the relevant part of the space of dimension-6 operators using an independent set of vertex corrections $\delta g$  and Wilson coefficients of 4-fermion operators. 
The latter are directly inherited from the Warsaw basis, such that the translation is trivial. 
The former are related to the Wilson coefficients of dimension-6 operators in the Warsaw basis by the following linear transformation: 
\bea
\label{eq:dgtowarsaw}
\delta g^{W e}_L & = &   c^{(3)}_{H \ell} + f(1/2,0) - f(-1/2,-1), 
%\nn
%\delta g^{Z \nu }_L & = &    {1 \over 2} c^{(3)}_{H\ell} - {1\over 2} c^{(1)}_{H\ell}   +  f(1/2,0),  
\nnl
\delta g^{Ze}_L & = &    - {1 \over 2} c^{(3)}_{H\ell} - {1\over 2} c_{H\ell}    +   f(-1/2, -1), 
\nnl
\delta g^{Ze}_R & = &    - {1\over 2} c_{He}   +  f(0, -1),  
\nnl 
%\eea
%\bea
%\delta g^{Wq}_L & = &   \left ( c^{(3)}_{H q}  + f(1/2,2/3) - f(-1/2,-1/3) \right ) V_{\rm CKM}  , 
%\nnl
\delta g^{Wq}_R & = &   - {1 \over 2 } c_{Hud},
\nnl 
\delta g^{Zu}_L & = &   {1 \over 2}  c^{(3)}_{Hq} - {1\over 2} c_{Hq}   + f(1/2,2/3), 
\nnl
\delta g^{Zd}_L & = &    -{1 \over 2}   V^\dagger  c^{(3)}_{Hq}   V   - {1\over 2} V^\dagger  c_{Hq}  V   + f(-1/2,-1/3),
\nn
\delta g^{Zu}_R & = &    - {1\over 2} c_{Hu}   +  f(0,2/3),  
\nn
\delta g^{Zd}_R & = &    - {1\over 2} c_{Hd}  +  f(0,-1/3), 
\eea 
where 
\bea
f(T^3,Q) &= & 
-   I_3   Q   {g_L g_Y \over g_L^2 - g_Y^2} c_{HWB} 
\\ &+ & 
 I_3 \left ( {1 \over 4 }[c_{\ell \ell}]_{1221}  - {1 \over 2}  [c^{(3)}_{H \ell } ]_{11}  -  {1 \over 2} [c^{(3)}_{H \ell } ]_{22} 
- {1 \over 4} c_{HD}  \right )  \left ( T^3 + Q {g_Y^2 \over g_L^2 - g_Y^2} \right ),  
\nonumber
\eea 
and $I_3$ is the $3 \times 3$ identity matrix in the generation space.  
Using \eref{dgtowarsaw} one can easily recast the results of this paper as a likelihood for the Wilson coefficients in the Warsaw basis. 
See Ref.~\cite{yr4} for the dictionary between $\delta g$ and the Wilson coefficients in the SILH basis.

%%%%%%%%%%%%%%%%%%%%%%%%%%%%%%%%%%%%%
\section{More general approach to low-energy flavor observables}
\label{app:LEFFE}
\setcounter{equation}{0}  

The low-energy flavor observables discussed in Ref.~\cite{Gonzalez-Alonso:2016etj} also probe precisely 4-fermion operators with a strange quark. 
In the framework of the SMEFT the corresponding observables receive contributions from flavor off-diagonal dimension-6 operators, and in this paper we marginalized our likelihood over them. 
We also approximated the CKM matrix as $V=1$ when acting on $\cO(\Lambda^{-2})$ terms in the Lagrangian.  
For completeness, in this appendix we provide the formalism that allows one to take into account the constraints from strange observables and retrieve the terms suppressed by off-diagonal elements of the CKM matrix. 
First, the effective low-energy Lagrangian in \eref{LE_nucc} is generalized to 
\bea 
\label{eq:LEFFE_nucc}
{\cal L}_{\rm eff} & \supset &
-\sum_{I,J=1,2} \frac{2 \tilde V_{uI}}{v^2} \left [  
\left ( 1+  \eL^{d_I e_J} \right ) (\bar e_J  \bar \sigma_\mu \nu_J)(\bar u \bar \sigma^\mu d_I) 
+ \eps_R^{d_I e}   (\bar e_J  \bar \sigma_\mu \nu_J)(u^c \sigma^\mu \bar d_I^c) 
\right . \nnl && \left . 
+ {\eps_S^{d_I e_J}  +\eps_P^{d_I e_J} \over 2} (e_J^c  \nu_J) (u^c d_I) 
+ {\eps_S^{d_I e_J}  - \eps_P^{d_I e_J} \over 2}  (e_J^c  \nu_J) (\bar u \bar d^c_I)  
\right . \nnl && \left . 
+ \eps_T^{d_I e_J}  (e_J^c  \sigma_{\mu \nu}  \nu_J) (u^c  \sigma_{\mu \nu} d_I) 
+ \hc  \right ], 
%\nnl
\eea 
such that it also includes charged currents with the strange quark ($s\to u\ell\nu_\ell$).  
At tree level, the low-energy parameters are related to the SMEFT parameters as 
\bea
\label{eq:epsilonL}
\epsilon_R^{d e}  =  - \eL^{d e} &= & \frac{1}{V_{ud}} \delta g_R^{W q_1} , 
\nnl  
\epsilon_R^{s e}  = - \eL^{s e} &=&  \frac{1}{V_{us}} [\delta g_R^{W q}]_{12} , 
\nnl 
 \eL^{d \mu} 
&= &  - {1 \over V_{ud}} \delta g_R^{W q_1} 
 +  \delta g_L^{W \mu}  -  \delta g_L^{W e} 
+  \left ( [c^{(3)}_{lq}]_{111 J} -  [c^{(3)}_{lq}]_{221 J}  \right ) {V_{Jd} \over V_{ud}}, 
\nnl 
 \eL^{s \mu}
&= & 
 - {1 \over V_{us}}[\delta g_R^{W q}]_{12}  
 +  \delta g_L^{W \mu}  -  \delta g_L^{W e}
+ \left (   [c^{(3)}_{lq}]_{111 J}   -  [c^{(3)}_{lq}]_{221 J}  \right ){V_{Js} \over V_{us}},
\eea  
\begin{eqnarray}
\label{eq:epsilonRSPT}
\epsilon_S^{d e_J}
&=& - \frac{1}{2 V_{ud}} \left ( V_{Kd} [c_{lequ}]^*_{JJ{K1}} +  [c_{ledq}]^{*}_{JJ11} \right ),
\nonumber\\
\epsilon_P^{d e_J}
&=& -  \frac{1}{2 V_{ud}} \left (  V_{Kd} [c_{lequ}]^*_{JJ{K1}} -  [c_{ledq}]^{*}_{JJ11} \right ),  
\nnl 
\epsilon_S^{s e_J}
&=& - \frac{1}{2 V_{us}} \left ( V_{Ks} [c_{lequ}]^*_{JJ{K1}} +  [c_{ledq}]^{*}_{JJ12} \right ),
\nonumber\\
\epsilon_P^{s e_J}
&=& -  \frac{1}{2 V_{us}} \left (  V_{Ks} [c_{lequ}]^*_{JJ{K1}} -  [c_{ledq}]^{*}_{JJ12} \right ),  
\nonumber\\
\epsilon_T^{d e_J}
&=& - \frac{V_{Kd}}{2 V_{ud}} [c^{(3)}_{lequ}]^*_{JJ{K1}}~,
\nonumber\\
\epsilon_T^{s e_J}
&=& - \frac{V_{Ks}}{2 V_{us}} [c^{(3)}_{lequ}]^*_{JJ{K1}}~.
\eea 
In addition to $\tilde V_{ud}$ we also introduce the  the rescaled CKM matrix element  parameter $\tilde V_{us}$.
Both are  distinct from the elements of the unitary matrix $V$, to which they are  related by
 $V_{ud} = \tilde V_{ud} (1 + \delta V_{ud})$, 
  $V_{us} = \tilde V_{us} (1 + \delta V_{us})$, where 
\bea
\delta V_{ud} & = &    - {1 \over V_{ud}} \delta g_L^{W q_1}  - \frac{1}{V_{ud}} \delta g_R^{W q_1}  +  \delta g_L^{W \mu} - {1\over 2} [c_{\ell \ell}]_{1221}   +  [c^{(3)}_{lq}]_{111 J} {V_{Jd} \over V_{ud}}, 
\nnl 
\delta V_{us} & = &  - {1 \over V_{us}} [\delta g_L^{W q}]_{12} -  \frac{1}{V_{us}} [\delta g_R^{W q}]_{12}  + \delta g_L^{W \mu}
- {1\over 2} [c_{\ell \ell}]_{1221}  +   [c^{(3)}_{lq}]_{111 J} {V_{Js} \over V_{us}}~.
\eea
The purpose of this rescaling is  to impose  the relation $\eL^{d_I e} =   - \eps_R^{d_I e}$ in \eref{LEFFE_nucc}.
After the rescaling, $\tilde V_{ud}$ and  $\tilde V_{us}$ are no longer related by the standard unitarity equation. 
In the limit where the mixing with the 3rd generation is neglected  we have  
$|\tilde V_{ud}|^2  + |\tilde V_{us}|^2 = 1 + \Delta_{\rm CKM} $,  where  
\bea 
\label{eq:deltackm1}
\Delta_{\rm CKM}  & =  & - 2 V_{ud }\delta V_{ud} - 2 V_{us}\delta V_{us} 
\nnl  & =  &
 2 V_{ud} \left (  \delta g_L^{W q_1}  +  \delta g_R^{W q_1}  -  [c^{(3)}_{lq}]_{111 J} V_{Jd} \right )
 + 2 V_{us}   \left (  [\delta g_L^{W q}]_{12}  + [\delta g_R^{W q}]_{12} -  [c^{(3)}_{lq}]_{111 J} V_{Js}  \right )
 \nnl &&- 
 2 \delta g_L^{W \mu} +  [c_{\ell \ell}]_{1221} . 
\eea 
As before, $\tilde V_{ud}$ may be affected by new physics contributing to $\eps_S^{d e}$ and should be treated as a free parameter in the fit. 
Ref.~\cite{Gonzalez-Alonso:2016etj} obtained the following constraints on the low-energy parameters 
\bea
\left(
\begin{array}{c}
\tilde{V}_{ud}^{e} \\
\Delta_{\rm CKM} \\
\Delta_L^s \\
\Delta^d_{LP}\\
  \epsilon_P^{de} \\
 \epsilon_R^{de}\\
 \epsilon_P^{se} \\
 \epsilon_P^{s\mu} \\
 \epsilon_R^s \\
 \epsilon_S^{s\mu} \\
 \epsilon_T^{s\mu} \\
 \epsilon_S^{de} \\
\eT^{de} \\
 \eS^{se} \\
 \eT^{se} \\
\end{array}
\right)
=
\left(
\begin{array}{c}
 0.97451\pm 0.00038 \\
 -1.2\pm 8.4 \\
 1.0\pm 2.5 \\
 1.9\pm 3.8 \\
 4.0\pm 7.8 \\
 -1.3\pm 1.7 \\
 -0.4\pm 2.1 \\
 -0.7\pm 4.3 \\
 0.1\pm 5.0 \\
 -3.9\pm 4.9 \\
 0.5\pm 5.2 \\
 1.4\pm 1.3 \\
 1.0\pm 8.0 \\
 -1.6\pm 3.3 \\
 0.9\pm 1.8 \\
\end{array}
\right)\times 10^{\wedge}\left(
\begin{array}{c}
 0 \\
 -4 \\
 -3 \\
 -2 \\
 -6 \\
 -2 \\
 -5 \\
 -3 \\
 -2 \\
 -4 \\
 -3 \\
 -3 \\
 -4 \\
 -3 \\
 -2 \\
\end{array}
\right)~,
\label{eq:LEFFEbounds}
\eea
in the $\overline{MS}$ scheme at $\mu=2$ GeV. Here $\Delta^s_{L} = \eL^{s\mu}- \eL^{se}$ and $\Delta^d_{LP} \approx \eL^{de}- \eL^{d\mu} + 24  \epsilon_P^{d\mu}$. The associated correlation matrix is given in Ref.~\cite{Gonzalez-Alonso:2016etj}. We note that some entries in this matrix are very close to one, so it is crucial to take it into account.

%%%%%%%%%%%%%%%%%%%%%%%%%%%%%%%%%%%%%%
\section*{Acknowledgements}

We thank Jens Erler, Jorge de Blas, and Miko{\l}aj Misiak  for enlightening discussions and correspondence. 
AF~is supported by the ERC Advanced Grant Higgs@LHC. M.G.-A. is grateful to the LABEX Lyon Institute of Origins (ANR-10-LABX-0066) of the Universit\'e de Lyon for its financial support within the program ANR-11-IDEX-0007 of the French government.
This research project has been supported by a Marie Sk{\l}odowska-Curie Individual Fellowship of the
European Commission's Horizon 2020 Programme under contract number 745954 Tau-SYNERGIES.

\bibliographystyle{JHEP}
\bibliography{llqqpaper.bib}

\end{document}